\documentclass[12pt,preprint]{aastex}

\usepackage{epsfig}

\usepackage{color}

\begin{document}

\title{Evolution and Distribution of Magnetic Fields from AGNs in Galaxy Clusters II. The Effects of
Cluster Size and Dynamical State}

\author{Hao Xu\altaffilmark{1}, 
Hui Li\altaffilmark{1,2},  
David C. Collins\altaffilmark{1,3},  
Shengtai Li\altaffilmark{1},
and Michael L. Norman\altaffilmark{3}}
\altaffiltext{1}{Theoretical Division, Los Alamos National Laboratory, Los
  Alamos, NM 87545; hao\_xu@lanl.gov, hli@lanl.gov, dccollins@lanl.gov, sli@lanl.gov} 
\altaffiltext{2}{Center for Magnetic Self-Organization in Laboratory and Astrophysical Plasmas}  
\altaffiltext{3}{Center for Astrophysics and Space Sciences,
  University of California, 
San Diego, 9500 Gilman Drive, La Jolla, CA 92093;  mlnorman@ucsd.edu}

\begin{abstract}
 Theory and simulations suggest that magnetic fields from radio jets and lobes powered by their central super massive black holes can be an important source of magnetic fields in the galaxy clusters. This is paper II in a series of studies where we present self-consistent high-resolution adaptive mesh refinement cosmological magnetohydrodynamic (MHD) simulations that simultaneously follow the formation of a galaxy cluster and evolution of magnetic fields ejected by an active galactic nucleus (AGN). We studied 12 different galaxy clusters with virial masses ranging from 1 $\times$ 10$^{14}$ to 2 $\times$ 10$^{15}$ M$_{\odot}$. In this work we examine the effects of the mass and merger history on the final magnetic properties. We find that the evolution of magnetic fields is qualitatively similar to those of previous studies. In most clusters, the injected magnetic fields can be transported throughout the cluster and be further amplified by the intra-cluster medium (ICM) turbulence during the cluster formation process with hierarchical mergers, while the amplification history and the magnetic field distribution depend on the cluster formation and magnetism history. This can be very different for different clusters.  The total magnetic energies in these clusters are between 4 $\times$ 10$^{57}$ and $10^{61}$ erg, which is mainly decided by the cluster mass, scaling approximately with the square of the total mass. Dynamically older relaxed clusters usually have more magnetic fields in their ICM. The dynamically very young clusters may be magnetized weakly since there is not enough time for magnetic fields to be amplified. 
  
\end{abstract}

\section{Introduction}

The increasing detections of radio emission from galaxy clusters, called radio halos and relics \citep[see][]{Carilli02, 
Ferrari08, Giovanini09} suggest that the ICM is permeated with magnetic fields. Radio halos 
are generally diffuse and extended over $\ge 1$ Mpc, covering the whole clusters, while radio 
relics, which are often observed at the edges at clusters, can extend as long as 2 Mpc \citep[e.g.][]{Weeren10}. 
By assuming that the total energy in relativistic electrons is comparable to the magnetic energy, the 
magnetic fields in the cluster halos can be estimated at $0.1-1.0$ $\mu$G and the total magnetic 
energy can be as high as $10^{61}$ erg \citep{Feretti99}.

Magnetic fields in galaxy clusters  are also extensively studied through Faraday rotation 
measurements (FRM). Distribution of FRM, combined with the ICM density measurement, often
yields cluster magnetic fields of a few to ten $\mu$G level, mostly in the cluster centers \citep{Carilli02}. 
More interestingly, FRM was investigated to suggest that cluster  magnetic fields may have a 
Kolmogorov-like turbulent spectrum in core regions \citep{Vogt05}, with an energy spectrum peak 
at several kpc. Other studies \citep[for example][]{Taylor93, Eilek02} have suggested that the coherence 
scales of magnetic fields can range from a few kpc to a few hundred kpc, implying large amounts of 
magnetic energy and fluxes in the ICM \citep{Colgate00}. Recently, study of magnetic fields  by 
FRM \citep{Govoni06,Guidetti08,Bonafede10} are extended to the outer part of clusters by 
observations of more radio galaxies behind or embedded in these clusters. It is expected that the 
Extened Very Large Array (EVLA) will provide unprecedented new observations 
on the magnetic fields in the ICM.

It is unlikely that magnetic fields have been dynamically important during the cluster formation. 
But, it is suggested that the strength and geometry of magnetic fields in clusters may play a crucial role 
in cluster formation through other processes, such as heat transport, 
which consequently affect the applicability of clusters as sensitive probes for cosmological
parameters \citep{Voit05}. In addition, since magnetic fields are closely related to synchrotron 
emission and FRM, the distribution of magnetic fields is important to the understanding the radio 
observations of the ICM.

Magnetic field evolution is highly nonlinear during cluster formation and difficult  to be studied analytically.  
 Cosmological MHD simulations are being used to study the properties of magnetic fields in the ICM.
 Such simulations can be very useful in interpreting the magnetic field information from the observation results. 
Since the origin of cluster magnetic fields is still being debated \citep{Widrow02},  various initial magnetic fields are
used in MHD cluster formation simulations. Simulations were done with initial magnetic fields 
either from some random or uniform fields at high redshifts \citep{Dolag02, Dubois08, Dubois09} 
or from the outflows of normal galaxies \citep{Donnert08}. 
All these simulations have found that fields can be further amplified by cluster merger
\citep{Roettiger99} and turbulence in addition to collapse, and their findings roughly match results from observations.
On the other hand, very small seed fields from some first principle mechanism, like Biermann battery effect, are 
also studied \citep{Kulsrud97,Xu09b} in galaxy cluster simulations. \citet{Kulsrud97} and \citet{Ryu08} suggested that very week 
seed fields can be amplified by dynamo processes in clusters, though current simulations 
have not been able to model such processes self-consistently.

Magnetic fields in large scale radio jets and lobes from AGNs serve as one very intriguing
source of cluster magnetic fields because observations show that they can carry large amounts of magnetic 
energy and flux \citep{Burbidge59,Kronberg01,Croston05, McNamara07}.  The magnetization
of the ICM and the wider inter-galactic medium (IGM) by AGNs has been
suggested on the energetic grounds
\citep{Colgate00, Furlanetto01, Kronberg01}, without details of the physical
processes of magnetic field transportation and amplification.  
Through cosmological MHD simulations, \citet{Xu09} showed that
magnetic fields injected  from a single AGN into a local region can be sufficient to magnetize
the whole cluster to the micro Gauss level by the operation of small-scale dynamo \citep{Brandenburg05,Subramanian06}.
Recently, \citet{Sur10} used MHD simulations to show that same process can also operate during the formation of first stars.
Small-scale dynamo may be  an important process to generate magnetic fields in various cosmic objects \citep{Schleicher10}. 

In the first paper \citep{Xu10} of this study, we studied the magnetic field evolution in a single massive, 
relaxed galaxy cluster with different AGN injected energies and injection redshifts.  That paper found that, 
as long as the magnetic fields are injected before the active merger period during the cluster 
formation history, the AGN magnetic fields can be spread throughout the whole cluster and get 
substantial amplification. The final magnetic fields are weakly dependent on the amount of initial 
magnetic fields. But the behavior of magnetic fields in clusters of different masses and at various 
dynamical states, young (unrelaxed) or old (relaxed) clusters, is still not well studied.

In this paper, we perform a series of high resolution adaptive mesh refinement (AMR) MHD 
simulations of 12 galaxy clusters of different masses with initial magnetic fields injected by 
an AGN.  This allows us to investigate the robustness of magnetizing the ICM using the AGN 
magnetic fields and address additional questions that are not examined with a single cluster. 
We explore the properties of magnetic field distribution and their evolution in the ICM of those 
clusters in different dynamical states during their formation histories. The organization of this 
paper is as follows. In Section \ref{sec:model}, we provide the details of the simulation setup, 
including the cluster formation and the magnetic fields injection. We then summarize the main 
results in Section \ref{sec:result}. We present the detailed spatial distribution of magnetic fields, 
evolution of the magnetic energy and the radial profiles of magnetic fields strength. We also 
present and discuss the properties of synthetic Faraday rotation measurement of our simulated 
clusters.  In Section 4, we present a summary of the main findings and conclusions.

\section{Basic Model and Simulations}
\label{sec:model}

We have performed self-consistent adaptive mesh refinement (AMR) cosmological MHD galaxy cluster formation simulations of a 
set of clusters with initial magnetic fields injected by an AGN, using cosmological
MHD code ENZO+MHD \citep{Collins09}. 
The simulation setup is the same as that in \citet{Xu10}. The initial conditions 
are generated from an \citet{Eisenstein99} power spectrum. The simulations use 
a $\Lambda$CDM model with parameters $h=0.73$, $\Omega_{m}=0.27$, 
$\Omega_{b}=0.044$, $\Omega_{\Lambda}=0.73$, and $\sigma_{8}=0.77$. These 
parameters are close to the values from recent WMAP 3 results \citep{Spergel07}.
Initial conditions from different random seeds are used to generate different clusters.
The simulation volume of each run is $256$ $h^{-1}$Mpc on a side, and it uses a $128^3$
root grid and $2$ level nested static grids in the Lagrangian region
where the cluster forms. This gives an effective root grid resolution
of $512^3$ cells ($\sim$ 0.69 Mpc) and dark matter particles of mass resolution of $1.07
\times 10^{10}M_{\odot}$. During simulations, $8$ levels of refinement are allowed beyond 
the root grid, for a maximum spatial resolution of $7.8125$ $h^{-1}$kpc.

The simulations are evolved from redshift $z=30$ to $z=0$ using an adiabatic equation of state, with gamma=5/3.  The 
Simulations do not include other physics, such as radiative 
cooling or star formation feedback, as they are not important to the majority of the dynamics of cluster formation. Here, we have simulations of  
12 galaxy clusters with mass ranging from 9.9 $\times$ 10$^{13}$ to 1.9 $\times$ 10$^{15}$ $M_\odot$ at $z=0$. 
In previous studies \citep{Xu10}, we have shown that the magnetic field evolution is neither sensitive to the
injection redshift between z=3 and z=1 nor to the injected AGN magnetic energy, so in this study we have only 
 one injected AGN magnetic energy and a single injection time. We ``turn on'' the AGN with magnetic field injection at redshift
$z=3$ in the most massive halo (or the second most massive one in some runs).  The initial magnetic fields are injected into the ICM locally assuming that they 
are from an AGN (see description in \citet{Xu08a,Xu09}), basing on a magnetic tower model \citep{Li06}. 
 The injected magnetic energy is $\sim$ 6 $\times$ 10$^{59}$ erg for all runs.
Based on the dynamical states of these simulated clusters,
these 12 clusters fall into 2 groups at z=0, each has 6  clusters. We consider the clusters, which have more than half of their final
masses by z=0.5, as relaxed clusters, and the clusters, which gain more than half of their final masses from z=0.5 to z=0, as unrelaxed clusters. 
Group one is the relaxed clusters. They are labeled as R1 to R6, in the order of their final masses. Group two is the unrelaxed clusters. They are labeled as
U1 to U6, also in the order of their final masses. For two of the unrelaxed clusters (U1 and U2) we have performed 
two simulations, with magnetic energy injected into different progenitor halos. They are labeled as U1a, U1b, U2a and U2b, respectively.
The reason that we have injections in different locations is because the choice of initial magnetized halo 
turns out to play an important role in the final magnetism of these two clusters. So we 
have a total of 14 simulations. The properties of the injection halos and the final clusters are summarized in Table \ref{table:halos}, 
while the injection magnetic energies are listed in Table \ref{table:energy}.  

\begin{table}
\caption{Properties of simulated galaxy clusters.}
\begin{center}
\label{table:halos}
\begin{tabular}{|c|c|c|c|c|c|c|}
\hline
\hline
     & \multicolumn{3}{c|}{Cluster Properties (z=0)} & \multicolumn{3}{c|}{Cluster Properties (z=3)}    \\
Name &  $r_{200}(Mpc)$ &  $M_{vir}(M_\odot)$ & $M_{gas}(M_\odot)$  &   $r_{200}(Mpc)$ &  $M_{vir}(M_\odot)$ & $M_{gas}(M_\odot)$  \\
\hline
R1      &    2.161 & 1.252e15 & 1.863e14   &   0.194  & 1.577e13 & 2.296e12          \\
R2      &    1.909  & 8.633e14 & 1.215e14  &   0.171   & 1.111e13 & 1.495e12       \\
R3    &    1.590 & 4.985e14 & 7.116e13  &   0.160  & 9.132e12 & 1.223e12      \\ 
R4      &    1.364 & 3.149e14 & 5.000e13  &   0.191  & 1.548e13 & 2.109e12   \\
R5    &    1.148 & 1.877e14 & 2.864e13  &   0.190  & 1.536e13 & 2.160e12       \\
R6    &    0.927 & 9.897e13 & 1.491e13  &   0.171 & 1.113e13 & 1.541e12        \\
U1a    &    2.498 & 1.934e15 & 2.661e14    &   0.191 & 1.552e13 & 2.278e12      \\
U1b    &    2.468 & 1.866e15 & 2.733e14  &   0.143   & 6.540e12 & 9.612e11          \\
U2a    &    1.743   & 6.572e14 & 8.697e13  &   0.089 & 1.578e12 & 1.940e11    \\
U2b    &    1.718 & 6.292e14 & 7.963e13  &   0.142  & 6.349e12 & 9.033e11    \\ 
U3     &    1.709 & 6.198e14 & 8.105e13  &   0.116   & 3.460e12 & 4.269e11       \\
U4    &    1.675 & 5.829e14 & 7.665e13  &   0.126  & 4.449e12 & 5.954e11      \\
U5    &    1.626 & 5.338e14 & 7.370e13  &   0.133  & 5.318e12 & 7.007e11    \\ 
U6      &    1.447 & 3.763e14 & 5.312e13  &   0.146  & 6.908e12 & 9.070e11         \\
\hline
\end{tabular}
\end{center}
\end{table}

The AMR settings are the same as in \citet{Xu10}. The AMR is applied only in a region 
of ($\sim$ 50 Mpc)$^3$ where the galaxy cluster forms. During the course of cluster 
formation before the magnetic fields are injected, the refinement is controlled 
by baryon and dark matter overdensity. After magnetic field
injection, all the regions where magnetic field strength is higher than 
$5 \times 10^{-8}$ G are refined to the highest level, with 7.81 h$^{-1}$kpc resolution.  
The data analysis in this paper was performed 
using yt\footnote{http://yt.enzotools.org} for Enzo \citep{Turk11}. 

There are limitations with our current simulations, which should be keep in mind 
when interpreting the results from these simulations. 
The highest resolution of $\sim$10 kpc, which already made these simulations very big and
very computationally expensive, is close to the characteristic 
scale of the ICM magnetic fields obtained from observations. So our simulations miss some small
scale features of the magnetic fields. Higher resolution simulations 
are underway. 
In  addition, the MHD treatment without any kinetic effects in our simulations are not enough to completely 
understand the magnetic properties in the ICM. In some situations, the difference introduced by kinetic effects 
may be dramatic. For example, when effects of anisotropic pressure is included
in a weakly-collisional plasma, growth rate of magnetic field strength may be much 
higher \citep{Schekochihin06, Schekochihin08}. Unfortunately, the microphysical processes are still impossible to be simulated 
self-consistently in cosmological simulations with current computational capability. 

\section{Results}
\label{sec:result}

\subsection{Formation of the Galaxy Clusters and Evolution of Magnetic Field Distribution}
\label{sec:clusterformation}

\begin{figure}
\begin{center}
\epsfig{file=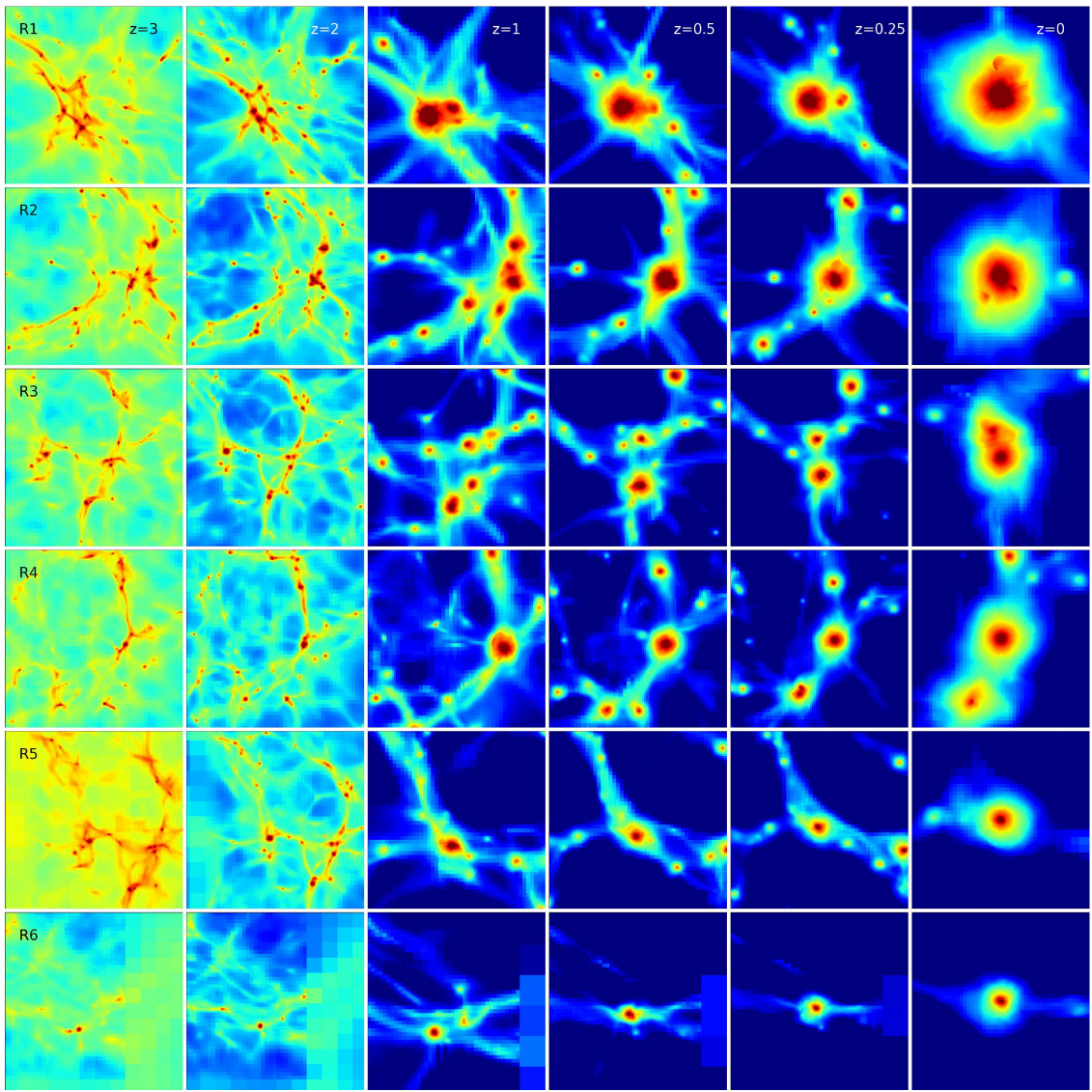,height=0.7\textheight}
\end{center}
\caption{Projected baryon density for relaxed clusters in group 1 at different redshifts. 
   The first column is projections at the time of injection, and each image covers a region of
  $32.0$ Mpc $\times$ $32.0$ Mpc comoving. The second column shows at z=2 and each image size is also 
  $32.0$ Mpc $\times$ $32.0$ Mpc comoving. The third, fourth and fifth columns are images at 
z=1, z=0.5, and z=0.25, respectively. The plot sizes are $16.0$ Mpc $\times$ $16.0$ Mpc comoving.
  The final column shows projections at z=0. Each image covers an area of $8.0$ Mpc $\times$ $8.0$ Mpc.  The first panel in each row is marked
  with the respective runs.  The color range of all plots is
  from $1.0 \times 10^{-4}$ to  $1.0 \times 10^{-2}$ g cm$^{-2}$.
\label{fig:density_r}}
\end{figure}

\begin{figure}
\begin{center}
\epsfig{file=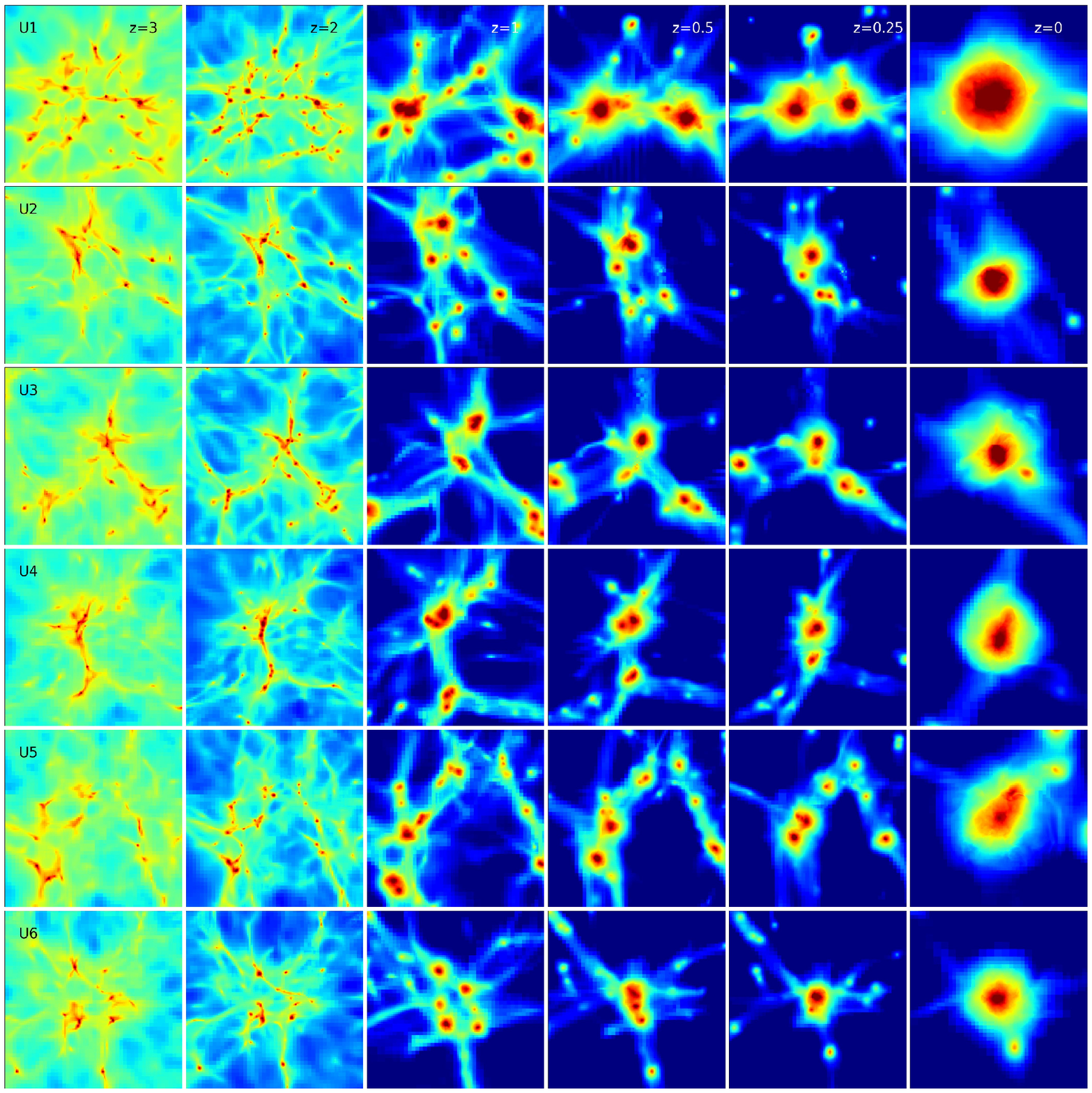,height=0.7\textheight}
\end{center}
\caption{ Same as Figure. \ref{fig:density_r}, but for unrelaxed clusters in group 2.
\label{fig:density_u}}
\end{figure}

We first briefly show the formation histories of our simulated clusters.  The hierarchy formation histories 
of our simulated clusters are presented by plotting the projected gas densities at various redshifts from 
z=3 to z=0. Relaxed clusters are plotted in Figure \ref{fig:density_r}, and 
the unrelaxed clusters are shown in Figure \ref{fig:density_u}. 
The clusters usually undergo numerous 
mergers at redshift between 3 and 0.5. 
For redshift less than $\sim$ 0.5, the clusters start to relax unless some rare big mergers happen.
The dynamically old and relaxed clusters in group 1 have experienced major mergers several 
Gyr before the end of the simulation, and have been dynamically relaxed for some time before z=0.  
In contrast, the younger, unrelaxed clusters of Group 2 have experienced major mergers closer to the 
end of the simulations. As we have mentioned at previous section,  the clusters in group 1 have more than half of their final
masses at z=0.5, while the clusters in group 2 more than double their masses from z=0.5 to z=0 by big mergers. 
From the shapes of the projected density, 
some clusters of group 2 look relaxed , e.g. cluster U1, though they have 
just finished a very big merger and their ICM motions are still quite active. The cluster U2 is undergoing 
active mergers after z=0.25 and is the dynamically youngest cluster in our simulations. 
This makes the evolution of magnetic fields in U2a and U2b very different from other runs.

\begin{figure}
\begin{center}
\epsfig{file=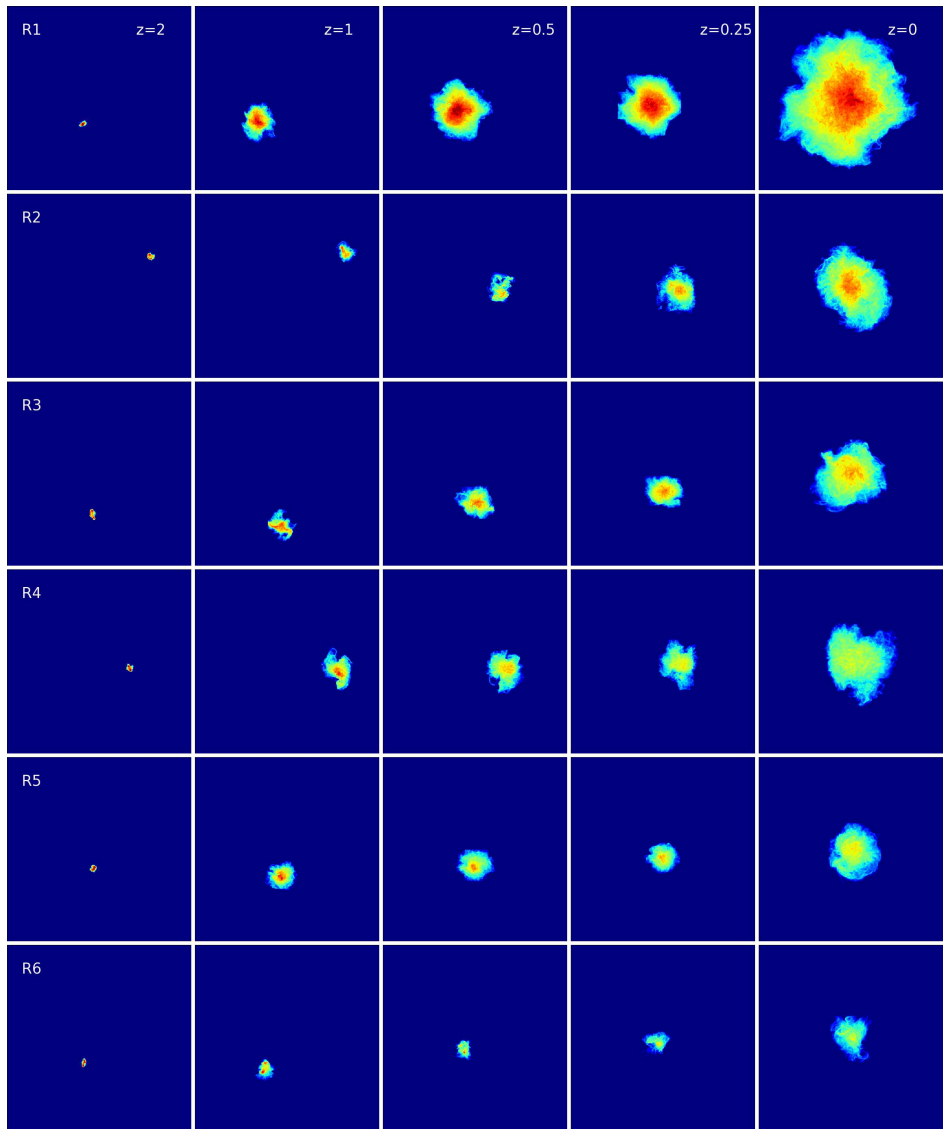,height=0.7\textheight} 
\end{center}
\caption{
Projected magnetic energy density for relaxed clusters at different redshifts. The first column shows 
 at z=2 and each image size is $32.0$ Mpc $\times$ $32.0$ Mpc comoving. The second, third and 
  third columns are images at z=1, z=0.5, and z=0.25, respectively. The plot sizes  are  
  $16.0$ Mpc $\times$ $16.0$ Mpc comoving. The final column shows projections at z=0. Each image 
   covers an area of $8.0$ Mpc $\times$ $8.0$ Mpc.  The first panel in each row is marked
  with the respective runs. The color range is from $1.0 \times 10^{8}$ to  $1.0 \times 10^{12}$
  erg cm$^{-2}$.
\label{fig:med_r}}
\end{figure}

\begin{figure}
\begin{center}
\epsfig{file=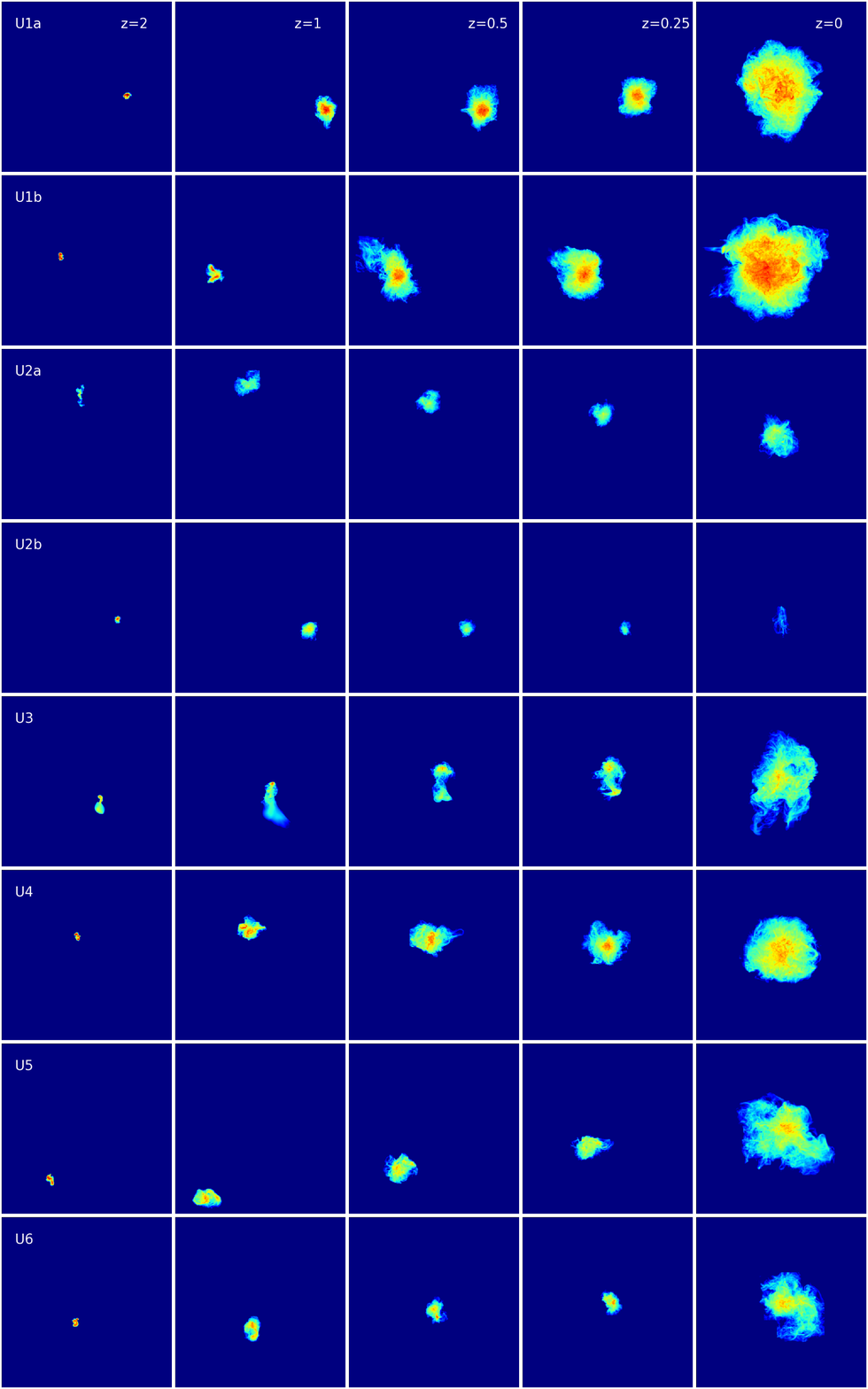,height=0.8\textheight} 
\end{center}
\caption{ Same as Figure. \ref{fig:med_r}, but for unrelaxed clusters in group 2.
\label{fig:med_u}}
\end{figure}

Figure \ref{fig:med_r} and \ref{fig:med_u} show the projected magnetic energy densities.
Magnetic fields mostly follow the motion of the halos in which magnetic fields are injected and may 
move a long distance during the  cluster formation. No matter where the magnetic fields are injected, 
the final magnetic fields tend to distribute around 
the cluster centers at z=0. These plots clearly show that mergers, especially major mergers,
spread the magnetic fields throughout the clusters. So the distribution of magnetic
fields are effected by when these big mergers happen during the cluster formation history.
There is a clear trend that the sizes of the magnetized areas and their associated magnetic energy densities 
are proportional to the sizes of their host halos. This is no surprise since larger clusters, which are formed 
by more mergers and have higher turbulence level to diffuse, amplify and maintain their magnetic fields.

Magnetic fields have already been spread to a large area of 
the whole clusters when simulations stop at $z=0$ in all runs, except for runs U2a and U2b. But the distribution of magnetic fields can be quite different from that
of the ICM gas if major cluster mergers have occurred recently. In such case, the magnetic field strength may peak away from 
the cluster centers. Magnetic fields in larger clusters, which are less  
affected by late time small mergers, are distributed more regularly.

In most runs, the halo with the initial magnetic injection grows fast and becomes the major one progenitor of the clusters 
during their formation histories, so their magnetic field evolution and distribution are similar within their groups.
But that is not true in three clusters, R2, U1 and U2.
For cluster R2, though it is well formed before $z=0.5$ and has relaxed at final output, 
its magnetized halo does't merge with the major one until z $\approx$ 1. So its magnetic fields  
can't spread to a large volume and get amplified in the early time.  This makes its magnetic fields
 weak and only locally distributed as a large relaxed cluster.
For the two unrelaxed clusters U1 and U2, they don't have a single major halo during their cluster formation until very
low redshift. Initial magnetic fields are then injected into two different halos to see the different magnetic filed evolution. 
The cluster U1 is finally formed by the merging of two similar size sub-clusters at z $\sim$ 0.2. In runs U1a and U1b, 
we initially magnetize one of these two sub-clusters, respectively. 
Their late major merger doesn't have enough time
to spread their magnetic fields to the other parts of the cluster. 
The cluster U2 is formed by active mergers  of several sub-clusters at very 
low redshifts. Runs U2a and U2b  have two of them magnetized initially. In both runs, the lack of big mergers 
in the early time causes the magnetic fields to have much less amplification,  
and their final magnetic fields are very weak and locally distributed at z=0. The magnetic fields
in run U2b are so weak that they are barely seen in Figure \ref{fig:med_u}

\subsection{Magnetic Energy Evolution}

\begin{figure}
\begin{center}
\epsfig{file=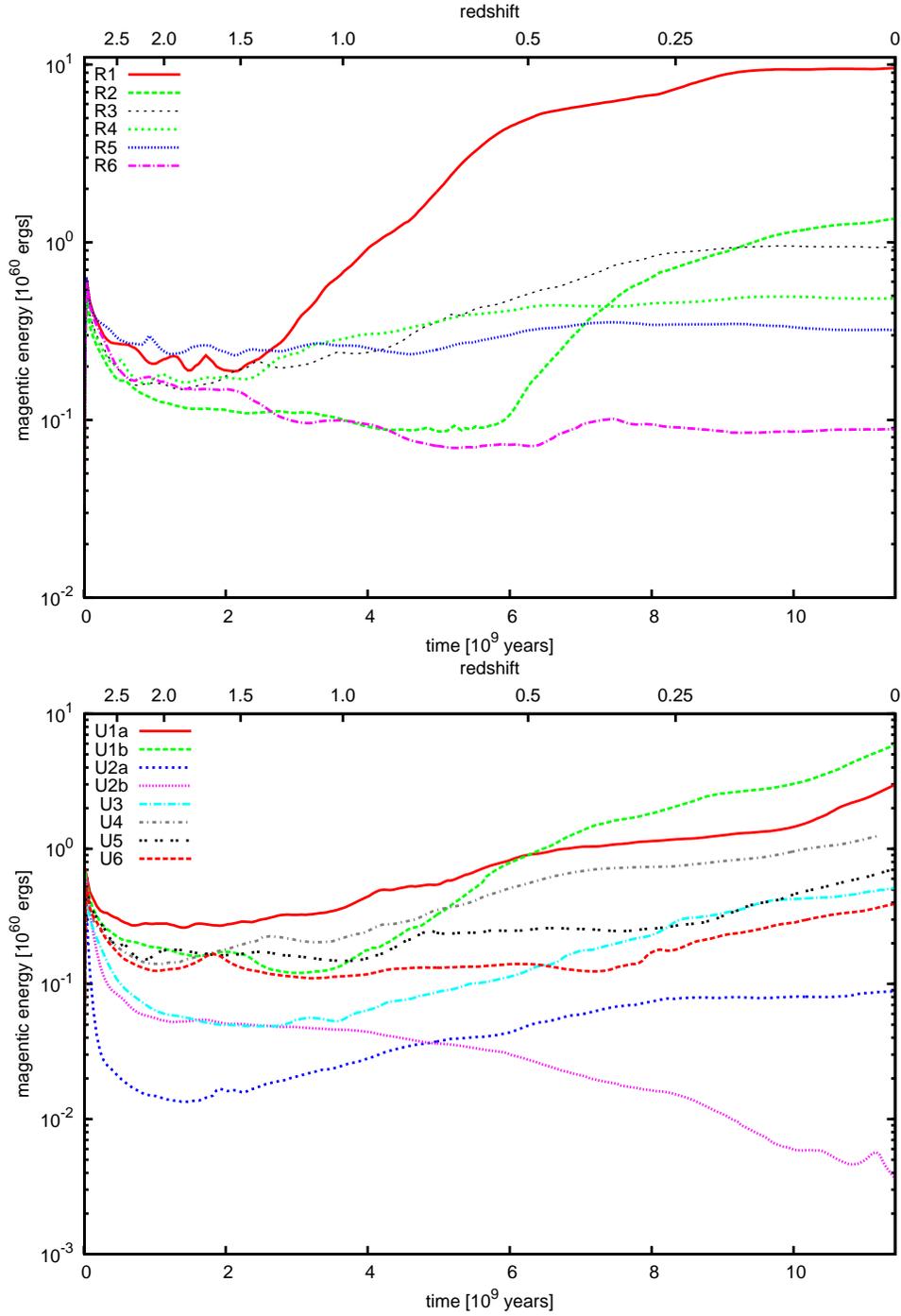,width=0.8\textwidth} 
\end{center}
\caption{Evolution of total magnetic energy of 
  the simulations, the upper panel shows relaxed clusters, while the
  bottom panel shows the unrelaxed clusters . 
\label{fig:energy}}
\end{figure}

We present the evolution of the total magnetic energy of all the simulations in Figure \ref{fig:energy}, separated into two groups of different dynamical states. 
The magnetic energy generally decreases rapidly initially, for a few hundred million years, because of the rapid expansion of the initial magnetic structure. It typically takes about 1-2 Gyr for the magnetic fields to expand in the ICM to a large volume to catch the turbulent motions and to be amplified.  Since we inject large amount magnetic energy into the system, which
may be much more than the magnetic energy that some small halos, like in cluster R6, 
can maintain. In such cases, their magnetic energies are keep decreasing
until they drop to the levels that their magnetic fields can be sustained in their host halos by the turbulence.  After that the magnetic energy gradually increases due to the increasing size and the ICM turbulent level by continuous mergers until saturation occurs (due to lack of mergers) or the simulation ends. Although the precise evolution of magnetic fields is quite different for various runs,
the fastest amplifications of magnetic fields generally happen a while (hundred million years) after the active mergers (big or continuous mergers)  which assemble the major part of the clusters. 

The final magnetic energies are between 
4 $\times$ 10$^{57}$ and 1 $\times$ 10$^{61}$ erg. The magnetic energy is 
saturated (stops growing) \citep[for a detailed discussion, see][]{Xu10} for several Gry for relaxed clusters in group 1. The only exception is cluster R2, whose weak magnetic fields, due to lack of early major mergers, 
are still being amplified long after the relaxation of the cluster.
Almost all of the clusters in group 2 have major mergers at low z, 
so their magnetic energies keep increasing until simulations end, except for run U2b. 
For this run, its magnetic fields, which are in a very small halo before the final assembly, 
occupy too small a volume to get amplified effectively when the simulation ends.

\begin{figure}
\begin{center}
\epsfig{file=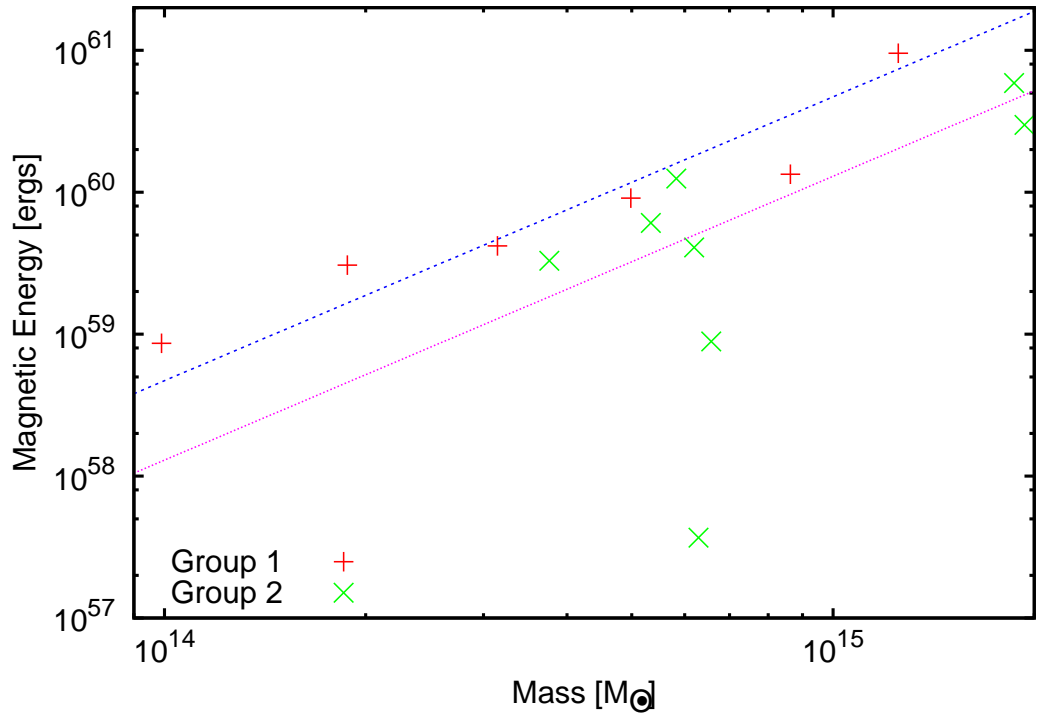,width=0.9\textwidth} 
\end{center}
\caption{Total magnetic energy vs. the virial mass of all simulations at z=0.
\label{fig:me_mass}}
\end{figure}

Not surprisingly, the total magnetic energy is higher for the larger clusters, which have deeper potential well, and higher thermal  
and kinetic energies.  We plot the total magnetic energy versus  the virial mass of the clusters in Figure  \ref{fig:me_mass} at $z=0$.  
The total magnetic energy is scaled as M$_{viral}^{2}$ for most of the runs, except runs U2a and U2b. 
The total magnetic energy and the virial mass can be a single proportionality  parameter $\alpha$ as :
\begin{equation}
\frac{E_M}{6.09 \times 10^{60} erg} = \alpha (\frac{M_{vir}}{10^{15} M_{\odot}})^2 
\label{equ:alpha}
\end{equation}
where $\alpha$ is within a range between 0.131 to 1.445 for all runs except runs U2a and U2b. The energy 6.09 $\times$ 10$^{60}$ is chosen to make $\alpha = 1$ for run R1. 
We summarize $\alpha$, magnetic energies and their ratio to kinetic energies inside R$_{200}$
of all simulations in Table \ref{table:energy}. 
The coefficient $\alpha$ is mostly decided by the cluster formation history. 
They are generally higher for the relaxed clusters in group 1 than for the dynamically younger clusters in group 2.
The magnetic energies of runs U2a and U2b of the youngest cluster are far away from the E$_{M}$ $\sim$
M$^{2}$ scaling. This is because their magnetic fields do not get amplified until the late active mergers, and their
current magnetic energies are limited by their small halos before the major active mergers. 

Though the processes of cluster formation and magnetic field evolution are highly nonlinear and complex,  
this scaling relation between magnetic energy and virial mass can be qualitatively understood. Using an overly-simplified isothermal model, the thermal
 and kinetic energies of galaxy clusters are proportional to M$_{vir}^{5/3}$ \citep{Bryan98} with a radial density profile as $\rho \propto r^{-2}$ 
 and a flat velocity dispersion. These relations were numerically confirmed by
 \citet{Vazza06, Vazza11} using both SPH and AMR codes. Since the clusters are not exactly isothermal, 
 especially in the outer part of the clusters,
 these scalings should be modified somewhat. 
 The profiles of the density and the velocity dispersion deviate from the isothermal distribution as
 $\rho \propto r^{-3}$ and $\sigma \propto r$ \citep{Navarro95, Sunyaev03} at large radii. 
 In such a case, we find that the  total kinetic energy is approximately proportional to
 the square of the virial masses in our simulations. As the magnetic energies in clusters in similar 
 dynamical state are proportional to the kinetic energies with
 small scattering related to formation and magnetization histories (since magnetic power spectra of all simulations have a same shape 
 and their levels are simply determined by the levels of the kinetic ones (see Section. \ref{sec:spectrum})), the magnetic 
 energy is then proportional to the square of the cluster mass. 

Since the magnetic energy in a halo is proportional to the square of it mass, the magnetic fields in a massive cluster are much larger
than the magnetic fields residing in its progenitor halos. So the major contributing factor to the final cluster magnetic fields is from the dynamo process after
majority of the cluster has formed. It also suggests that additional magnetic fields from more AGNs or 
regular galaxies in smaller halos may have only small impact on the final magnetic fields in a cluster.

\begin{figure}
\begin{center}
\epsfig{file=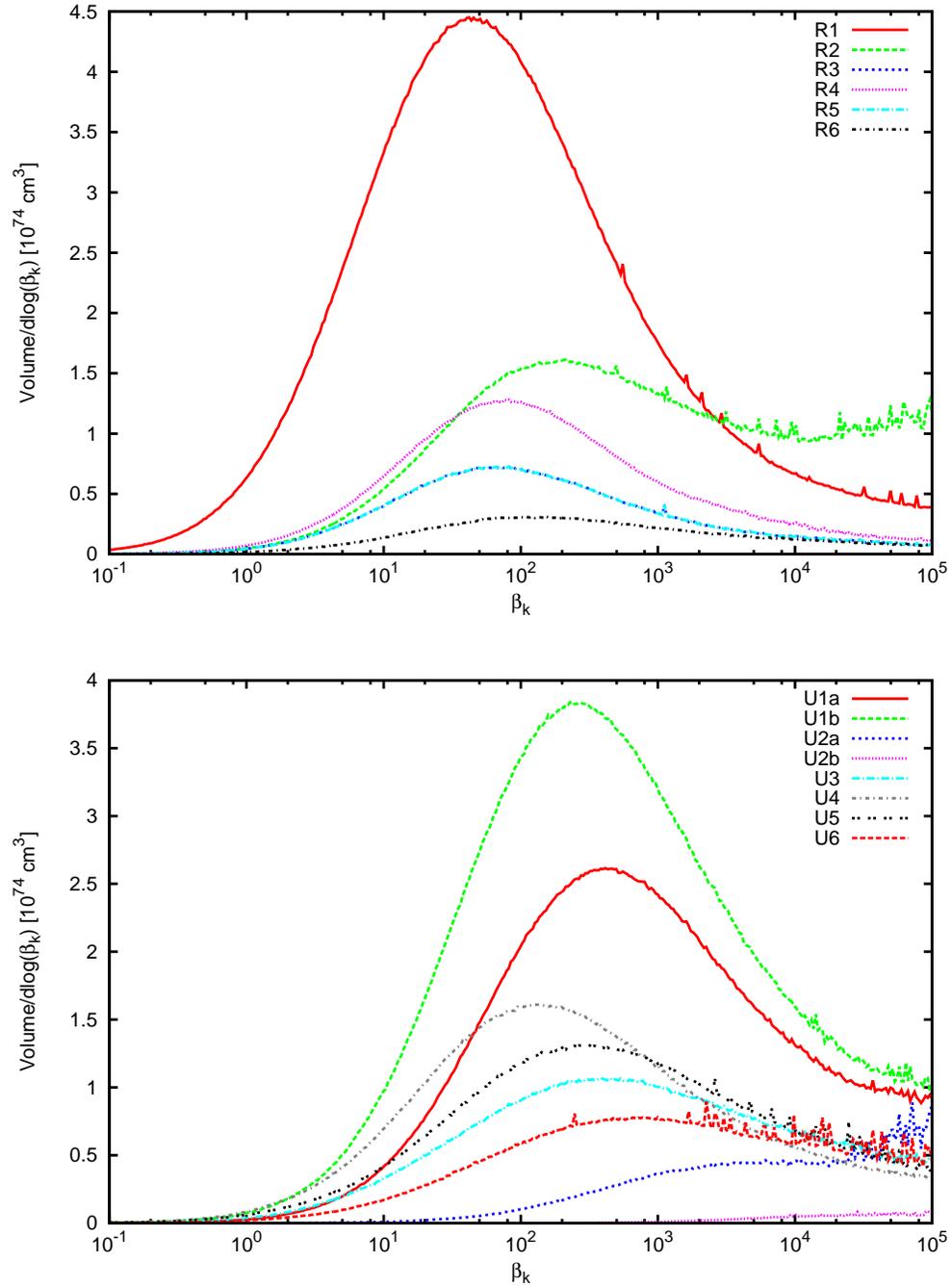,width=0.8\textwidth} 
\end{center}
\caption{Volume histogram of kinetic plasma $\beta$ (= e$_{K}$/e$_{M}$)
 inside the clusters' virial radii at z = 0. Relax clusters are shown in the top panel and
  the unrelaxed clusters are shown in the bottom panel. 
\label{fig:beta}}
\end{figure}

As listed in Table \ref{table:energy}, the magnetic energy in a cluster is only a small fraction of its kinetic energy. 
For relaxed clusters, their magnetic energies are about 1\% of their kinetic energies inside their virial radii. These ratios drop
to about 0.1\% for the dynamical younger clusters, and even smaller for the recently formed cluster U2. The local magnetic energy 
is much smaller than the kinetic energy in most of the cluster regions as well. To show this, we plot the volume 
histogram of kinetic $\beta$ (e$_{K}$/e$_{M}$) of all clusters at z=0 in 
Figure \ref{fig:beta}. For relaxed clusters, kinetic $\beta$ peaks at between 50 and 100, while peaks of 
kinetic $\beta$ are bigger than 100 for unrelaxed clusters. Even 
for a relaxed cluster, only a small fraction of its volume has magnetic energy bigger than kinetic energy.
So, the magnetic fields are dynamically unimportant in all these simulated clusters.
Several factors may contribute to why the magnetic energy is only a small fraction of the corresponding 
kinetic energy. First, it is possible that simulation resolution is not high enough, as the 
corresponding numerical diffusion is big, so the turbulence decays before
it has an opportunity to amplify the magnetic fields. Some of our higher resolution 
test runs show that there are somewhat larger amount of more magnetic fields, but 
the total magnetic energy is still much smaller than the kinetic energy. Second, it could be that 
there is not enough time during cluster formation for the magnetic fields to grow in large scales, 
of which the eddy turnover time is $\sim$ Gyr. Third, the driving
from mergers is not constant and not homogeneous, and becomes weaker 
when the cluster gets bigger. Fourth, it is not known what the expected level 
of magnetic field saturation is for compressible turbulence. Simulations of 
supersonic and superalfvenic compressible MHD turbulence in \citet{Kritsuk09} found that 
the saturated magnetic energy is much smaller than the kinetic energy if the initial fields are weak. 
A further study to understand why the magnetic fields saturate at a level much lower than the kinetic energy in the ICM is needed.
Fifth, since we only have one seed source for  magnetic fields, significant magnetic fields only fill a portion of the ICM 
(from 20\% to 80\% see later section on magnetic fields distribution). So in a case of more sources and
a higher filling of magnetic fields in a galaxy cluster, its magnetic energy ratio could potentially be 
somewhat higher (but still much smaller that kinetic energy, at most a few percent).

\begin{table}
\caption{Properties of magnetic energy in simulated galaxy clusters}
\begin{center}
\label{table:energy}
\begin{tabular}{|c|c|c|c|c|c|}
\hline
\hline
Name   & E$_M^{i}$ (erg)$^a$  & E$_M^{200}$ (erg)$^b$&  $\alpha$$^c$ &  $E_M^{200}/E_K^{200}$$^d$ & $\gamma$$^e$   \\
\hline
R1   &  6.17e59  & 9.55e60       & 1.00 &  0.00814  &  0.44 \\
R2   &  4.85e59 &   1.34e60      &  0.295 &  0.00306  & 0.65  \\
R3  &   5.57e59   &    9.10e59    &  0.601 &  0.00240 & 0.57 \\ 
R4  &   5.32e59   &   4.19e59     &  0.693 &  0.0122      & 0.3 \\
R5 &  6.37e59     &   3.07e59    &  1.436 & 0.0113   &  0.46 \\
R6   & 5.94e59      &   8.62e58     &  1.445 & 0.0113 & 0.45  \\
U1a         & 6.32e59      &   2.99e60  & 0.131 &  0.00119  & 0.56 \\
U1b         &   5.83e59    &   5.88e60    & 0.277 &  0.00248 &  0.55 \\
U2a         &  3.32e59     &    8.90e58    &  3.38e-2 &  1.77e-4 &  0.67 \\
U2b         &  6.01e59     &   3.68e57      &  1.53e-3 & 7.19e-6   & 0.81  \\ 
U3        & 4.74e59      &   4.09e59       & 0.175 & 8.21e-4  & 0.43 \\
U4        &  5.35e59     &   1.25e60        &  0.603 & 0.00513 & 0.74 \\
U5          &      5.35e59  &    6.07e59  &  0.350  & 0.00224 & 0.59 \\ 
U6        &   5.21e59     &  3.29e59    &  0.382  & 0.00259 & 0.53 \\

\hline
\end{tabular}
\end{center}
a: magnetic energy at the end of injection;               \\
b: magnetic energy inside R$_{200}$ at z=0;  \\ 
c: $\alpha$ in Equation \ref{equ:alpha}, proportionality  parameter between magnetic energy and cluster mass square;   \\
d: ratio of magnetic to kinetic energy inside R$_{200}$ at z=0;  \\
e:  best-fit of $\gamma$ in relation between the magnetic field strength and the gas density (B $\propto$ n$^{\gamma}$);
\end{table}

\subsection{Magnetic Field Distribution over the ICM at $z=0$}

\subsubsection{Radial Profiles of Magnetic Fields }
\label{sec:profile}

\begin{figure}
\begin{center}
\epsfig{file=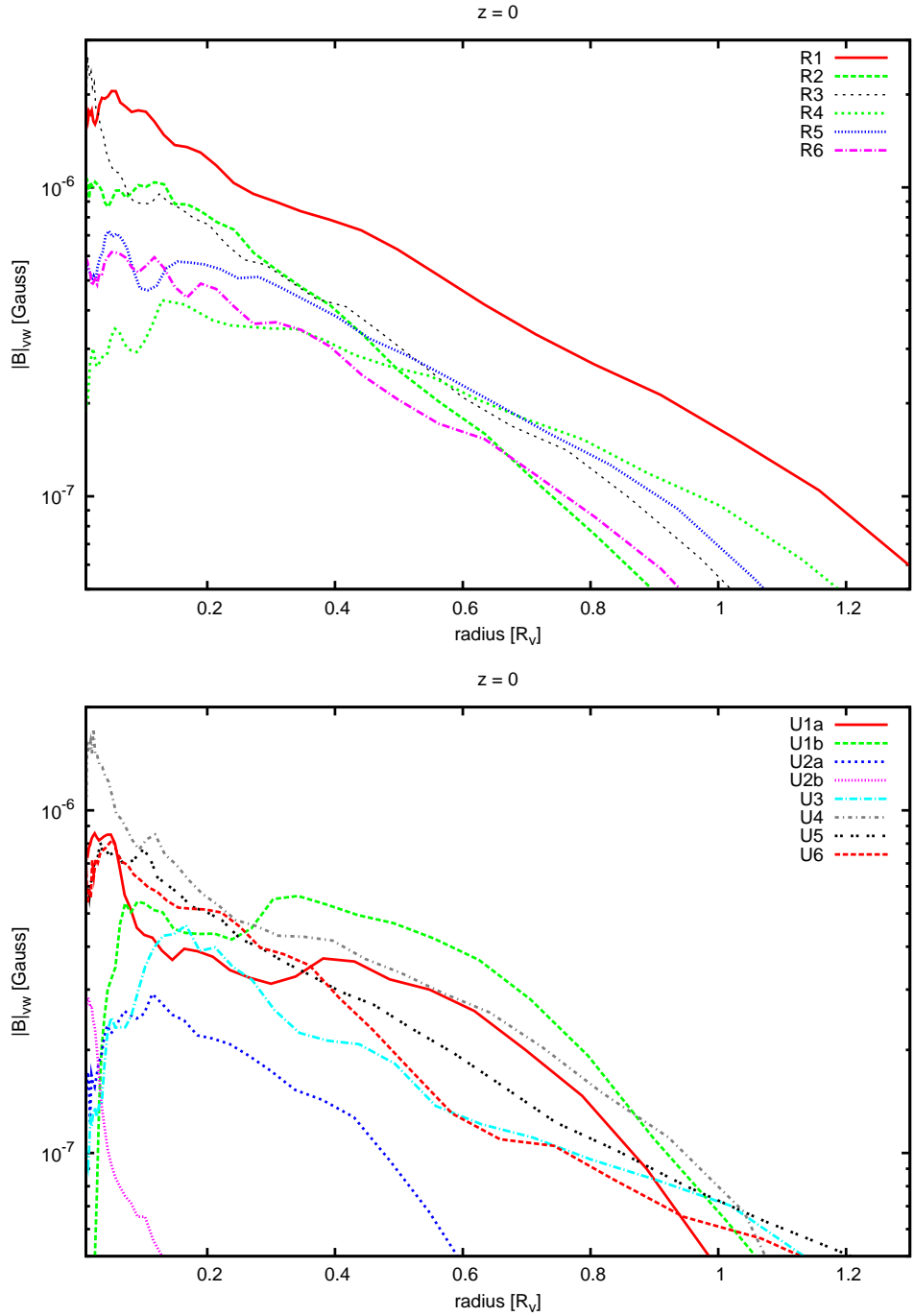,width=0.8\textwidth} 
\end{center}
\caption{Spherically averaged radial profiles of rms magnetic field strength of different
  clusters. Relax clusters are shown in the upper panel and the unrelaxed clusters are shown in the bottom panel.  The x-axis 
  is normalized by their virial radii.
\label{fig:Bprofile}}
\end{figure}

In Figure \ref{fig:Bprofile}, we present the spherically rms averaged radial
profiles of magnetic fields at z=0. The relaxed and unrelaxed clusters are 
plotted in two different panels, and the radii are normalized by their virial radii. 
It is clear that the profiles of magnetic field strength are determined both by the cluster 
sizes and their dynamical states. 

The magnetic field strength is generally higher for the bigger clusters and/or dynamically older clusters. For relaxed cluster, micro Gauss fields are present 
at the centers of clusters bigger that 10$^{15}$ M$_\odot$, but fields drop to about 0.5 $\mu$G for the 
smallest cluster of 10$^{14}$ M$_\odot$. The magnetic field profiles of relaxed clusters in group 1 are more regular as a function of radius,  while the magnetic field profiles of dynamically younger clusters in group 2 are obviously disturbed by their recent big mergers. 
For runs U2a and U2b, since the cluster is formed so late and no significant amplification of magnetic fields has been operated, 
their magnetic fields are much weaker and reside only in the inner part of the cluster.

 Several other MHD simulations \citep[e.g.][]{Dolag02,Dubois09}, as well as the study on the RM 
 and the X-ray surface brightness correlation \citep{Dolag01}, suggest
that the decline of magnetic fields is correlated with the gas density as n$^{\gamma}$. 
Analysis of RM and X-ray brightness data in \citet{Dolag01} 
yields the $\gamma = 0.9$ for A119, and $\gamma=0.5$ for 3C129 with large uncertainty. We 
fit our magnetic field radial profile with the gas density profile and list the best-fitted results in Table \ref{table:energy} .
The $\gamma$ in our simulations scatters between 0.3 and 0.81,  which is consistent with the results from \citet{Dolag01}. 
Relaxed clusters usually have smaller $\gamma$, so have flatter magnetic field profiles as they have more time 
to amplify their magnetic fields in the outer part by the ICM turbulence after major mergers. 
However, the reliability of our results may be limited by the way we magnetized clusters, we need simulations with 
many AGNs in one cluster to get more reliable n$^{\gamma}$ profiles. In addition, since this result is obtained by fitting the averaged radial
profiles of the magnetic field strength and the gas electron density, it doesn't mean that
there is a simple correlation between the magnetic field strength and the gas density throughout the cluster.

\subsubsection{Magnetic Field Spatial Distributions}

\begin{figure}
\begin{center}
\epsfig{file=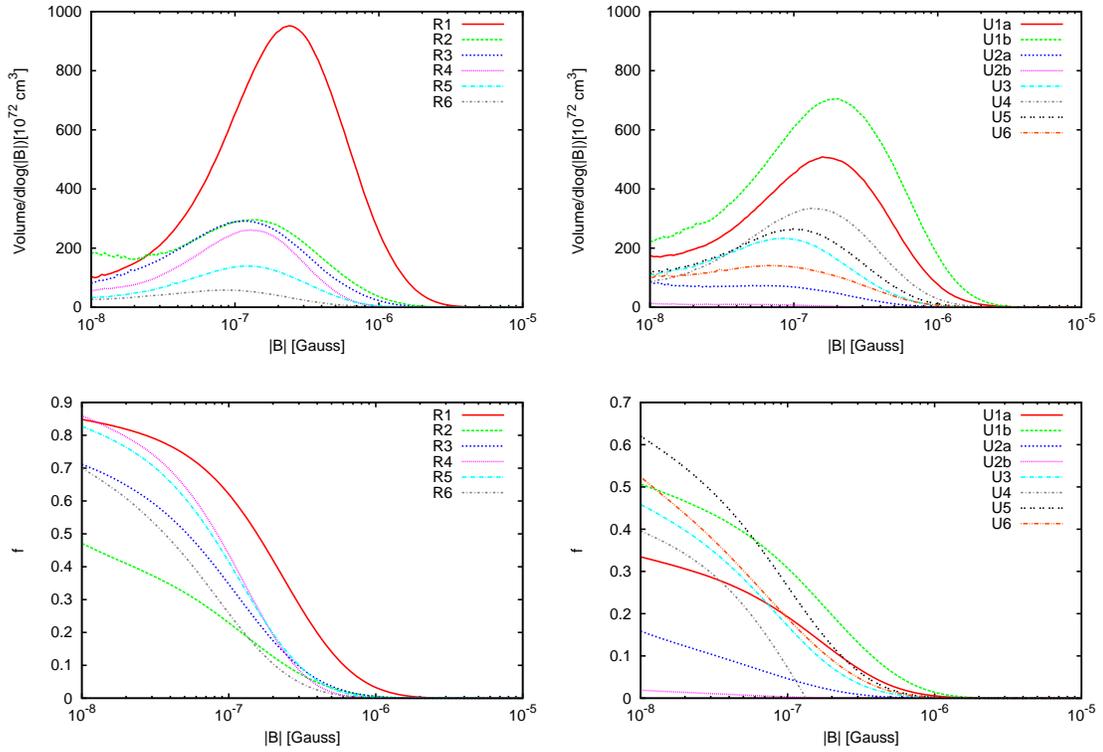,width=0.9\textwidth}
\end{center}
\caption{
Volume distribution (top) and complementary cumulative volume distribution (bottom) of the magnetic field strength 
inside the virial radius. 
The left panel shows the relaxed clusters, while the right panel shows the unrelaxed clusters.
\label{fig:filling}}
\end{figure}

We plot the volume histograms 
and complementary cumulative volume ratio histograms of magnetic field strength inside 
the virial radii for all simulations at z=0 in Fig. \ref{fig:filling}. The magnetic field fillings are quite different for
the relax and unrelaxed clusters. For dynamically older clusters, most volumes inside the virial radii are filled with magnetic
fields.  Magnetic fields stronger than 10$^{-8}$ G typically fill more than 75\% of the cluster volume for group one except run R2,  
and less than 50\% for group two except run U5. For run R2 of group one, since the magnetized halo merges with the bigger halo after z=1, 
its magnetic field volume filling is consequently smaller. 
Once the magnetic fields are well spread throughout the 
clusters and get amplified, magnetic field strength distribution peaks between 0.1 and 0.3 $\mu$G. 
Larger and older clusters have much higher peaks 
as well as longer tails of stronger magnetic fields. Runs U2a and U2b  
have very weak magnetic fields that only fill very small volumes. 

\begin{figure}
\begin{center}
\epsfig{file=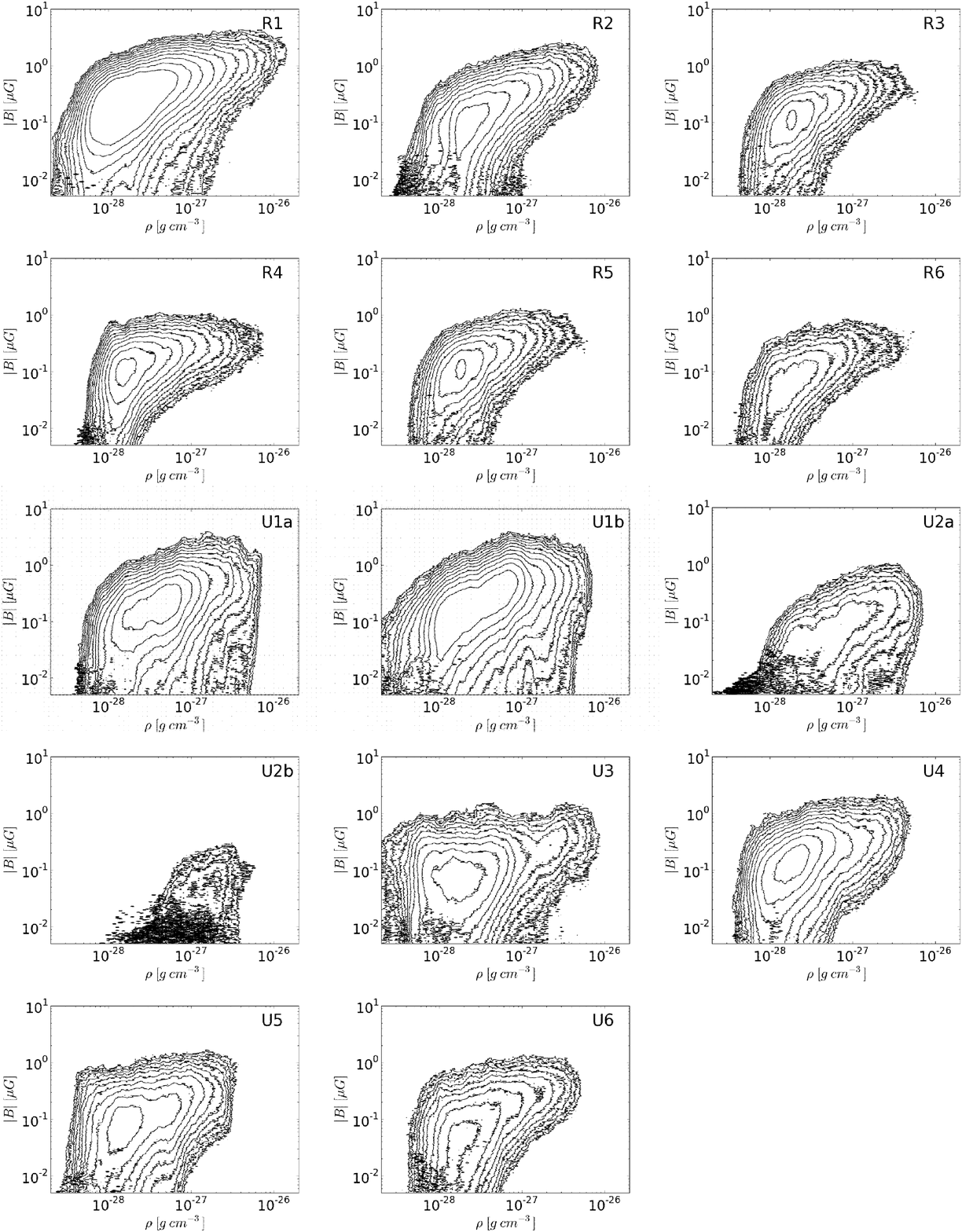,width=0.8\textwidth}
\end{center}
\caption{Volume weighted two-dimensional distributions of the magnetic field strength vs. baryon density of all runs inside the cluster virial
radiuses at z = 0. Contour lines are the volume of gas at that density and magnetic field at 10$^k$ cm3, where k = 69.0, 69.2, 69.4, . . . 71.2 from outer to inner.
The upper two rows are relaxed clusters, and the rest are unrelaxed clusters.}
\label{fig:contour}
\end{figure}

We also plot the two-dimensional distribution of the magnetic field strength 
verses the gas density inside their virial radii in Figure \ref{fig:contour},
showing how magnetic fields distributed over the ICM plasma.
There is no obvious correlation between the field strength and the gas density. The distribution is similar 
for clusters of similar sizes. Most of the magnetic field strength is between 0.1 to
a few micro Gauss, and are mixed with gas  over a wide range of densities.  This casts doubts on cluster magnetic field
modeling when simple correlation between $|B|$ and ICM density is assumed.

\subsection{Kinetic and Magnetic Energy Density Power Spectra}
\label{sec:spectrum}
 
\begin{figure}
\begin{center}
\epsfig{file=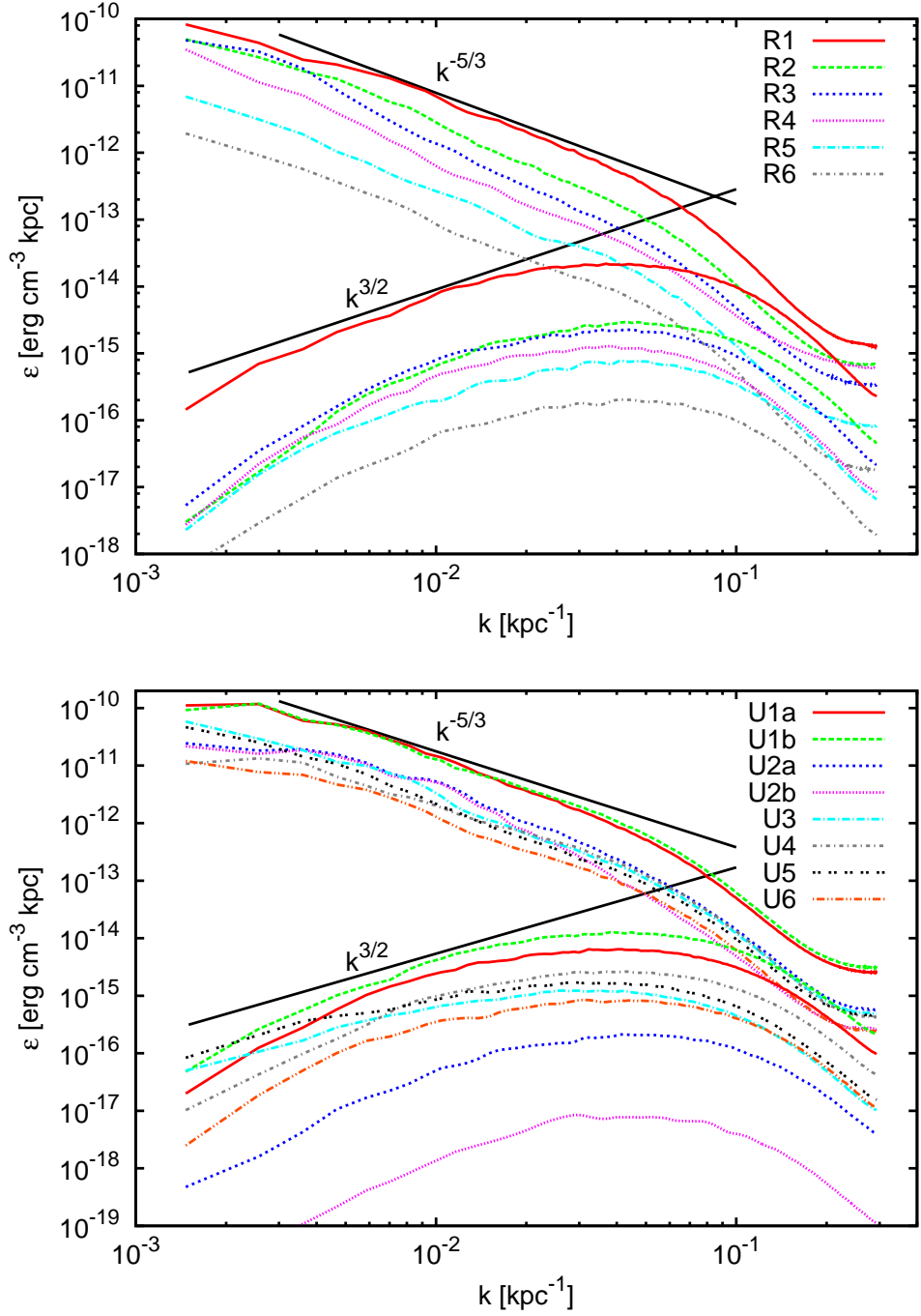,width=0.8\textwidth}
\end{center}
\caption{Kinetic and magnetic energy density spectra of relaxed (Top) and unrelaxed (Bottom) clusters. Kinetic and magnetic 
energy densities of same cluster are represented by same line. Kinetic energy densities of all clusters show k$^{-\frac{5}{3}}$ 
Kolmogorov-like spectra. All magnetic energy densities have a short k$^{\frac{3}{2}}$ Kazantsev spectrum, while the levels of the 
magnetic energy are depend on the cluster dynamical states and their magnetized histories.      }
\label{fig:spectra}
\end{figure}

The kinetic and magnetic energy density power spectra of all clusters at z=0 are shown in Figure \ref{fig:spectra}, 
which are computed from boxes of 512$^3$ cells ($\sim$(5.5 Mpc)$^3$) in the highest level enclosing the clusters. The ICM turbulence is represented 
by Kolmogorov-like spectra in kinetic energy. These kinetic spectra are also seen from pure hydrodynamics simulations of galaxy clusters \citep{Vazza09}.  
All magnetic spectra are in a similar shape and follow the k$^{3/2}$ Kazantsev law in the large scales.
The Kazantsev spectrum, which is the prediction of small-scale dynamo theory in the kinematic regime \citep[see][]{Brandenburg05}, 
is also found in simulations of a collapsing Bonnor-Ebert sphere in \citet{Federrath11}.
These magnetic spectra in our simulations show that the small-scale dynamo \citep[see][for detailed discussions]{Xu10} 
does operate in all our simulated clusters. For dynamically young clusters, some (like cluster U4 )
of their power spectra are disturbed by the recently mergers.  

For the relaxed or dynamically older clusters, the magnetic fields have enough time to be amplified, 
their magnetic energy densities are close to the kinetic energy densities in small scales. One possible
 reason that the magnetic energies in smaller scales are smaller than the
kinetic energies is because the magnetic fields do not fill all the space of the clusters. 
For the recently merged clusters,
the magnetic fields do not have enough time to spread through out the newly formed clusters and be completely amplified by
the ICM turbulence. So their magnetic energies are much smaller than the kinetic ones even in the small scales.

\subsection{Faraday Rotation Measurement from Simulations}

\begin{figure}
\begin{center}
\epsfig{file=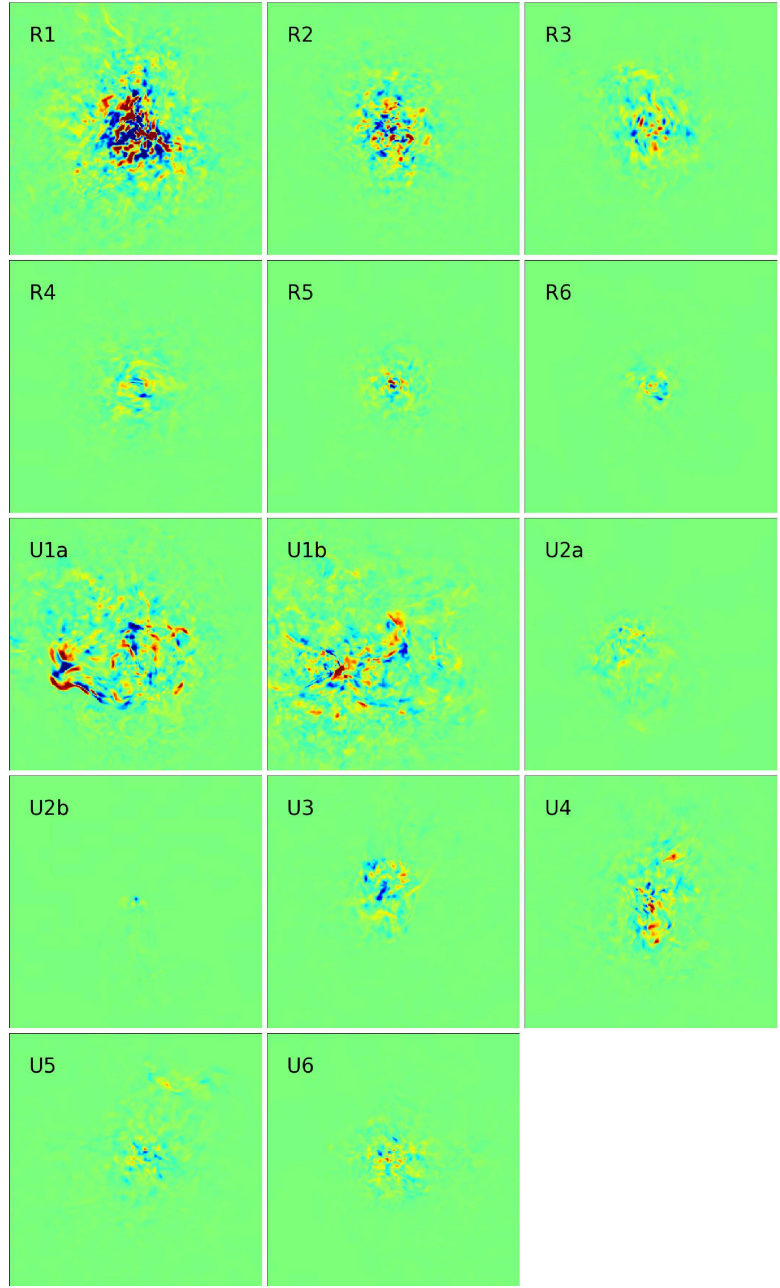,height=0.8\textheight} 
\end{center}
\caption{Faraday rotation measurement of the clusters by integrating
  through the cluster on the y direction. It covers a region of $3$ Mpc  
  $\times$ $3$ Mpc at z = 0.  The color range shown is from $-500$
  (blue) to $500$ (red) rad m$^{-2}$. The upper two rows are relaxed clusters, and the rest are unrelaxed clusters.
    \label{fig:rm}}
\end{figure}

The Faraday Rotation Measurement is a key method for measuring cosmic magnetic fields and has provided important information on cluster magnetic fields. 
The RM maps have been used 
not only to estimate the strength of magnetic fields (see review by \citet{Carilli02})
but also the turbulent structure of the fields \citep{Vogt05, Ensslin06} in the ICM.
We compute the RM maps using the magnetic field and free electron density distribution from our simulations by integrating
8 Mpc over the clusters centered at each cluster's center along the y axis at z=0.
We show our synthetic RM maps in Figure \ref{fig:rm}. Though the basic features of the RM distribution 
are similar to those from  previous study \citep{Xu09, Xu10},
clusters of different size and/or at different dynamical states have quite different RM distribution in their  absolute values,
cover areas, and structure scales. The morphologies of the RM maps reflect the dynamical states of the clusters and their merger histories.  
There is a clear trend that there are more small scale patterns in the 
dynamically old relaxed clusters than in young clusters. This is because they have  
more time for the magnetic field amplification in small scales after major mergers. On the other hand, 
there are more long filaments on the younger, unrelaxed clusters reflecting their recent mergers. There are very long filaments 
($>$ 1 Mpc) in runs U1a and U1b associated with their recent large head-on mergers.   
  It is interesting to see that the RM maps are quite different between these two runs, of which the pre-merger magnetic
fields reside in different sub-clusters, though the gas dynamics of these two runs are almost identical. 
This shows that the magnetism history of the cluster also plays an important role in determining their RM distribution. 

\begin{figure}
\begin{center}
\epsfig{file=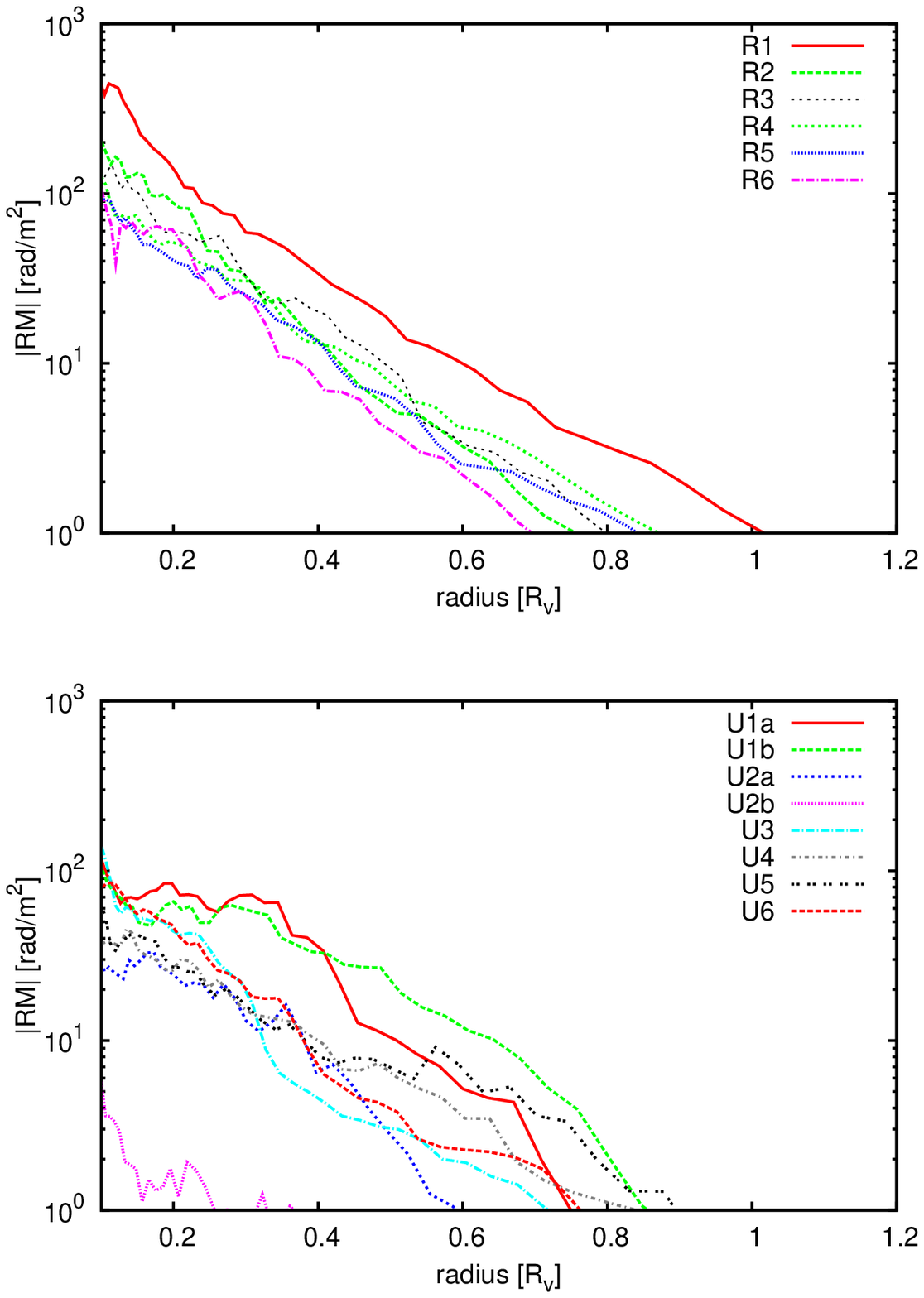,width=0.8\textwidth}
\end{center}
\caption{Azimuthally averaged radial profiles of $|RM|$ of the RM maps shown in Figure \ref{fig:rm}. 
The top panel shows the relaxed clusters, 
while the bottom panel shows that unrelaxed clusters.The x-axis 
  is normalized by the clusters' virial radii.}
\label{fig:rm_profile}
\end{figure}

\begin{figure}
\begin{center}
\epsfig{file=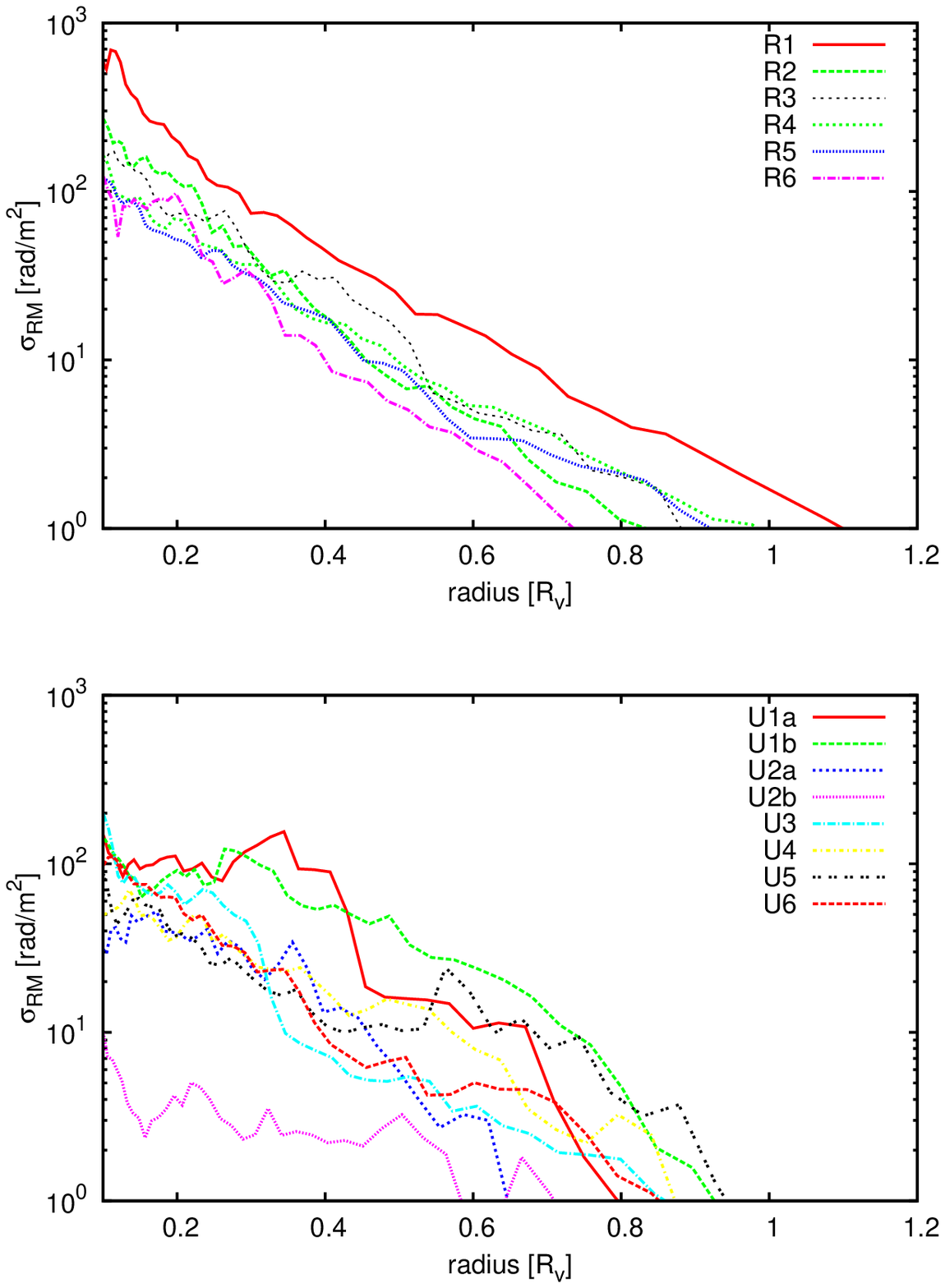,width=0.8\textwidth}
\end{center}
\caption{Azimuthally averaged radial profiles of standard deviation of $RM$ of the RM maps shown in Figure \ref{fig:rm}. 
The top panel shows the relaxed clusters, 
while the bottom panel shows that unrelaxed clusters.The x-axis 
  is normalized by the clusters' virial radii.}
\label{fig:sigma_rm_profile}
\end{figure}

The 2-D azimuthally averaged radial profiles of the absolute values of these FRM ($|RM|$)
are plotted in Figure \ref{fig:rm_profile}, while the radial profiles of the standard deviation of the RM are shown
in Figure \ref{fig:sigma_rm_profile}. The $|RM|$ profiles resemble the magnetic field strength 
profiles with steep slopes, as the ICM gas density decreases with the radius. These $|RM|$ profiles are 
roughly similar to the pattern from observational data in \citet{Clarke01}.  The $\sigma_{RM}$
profiles are also consistent with the recently observational results in  \citet{Govoni10}

\begin{figure}
\begin{center}
\epsfig{file=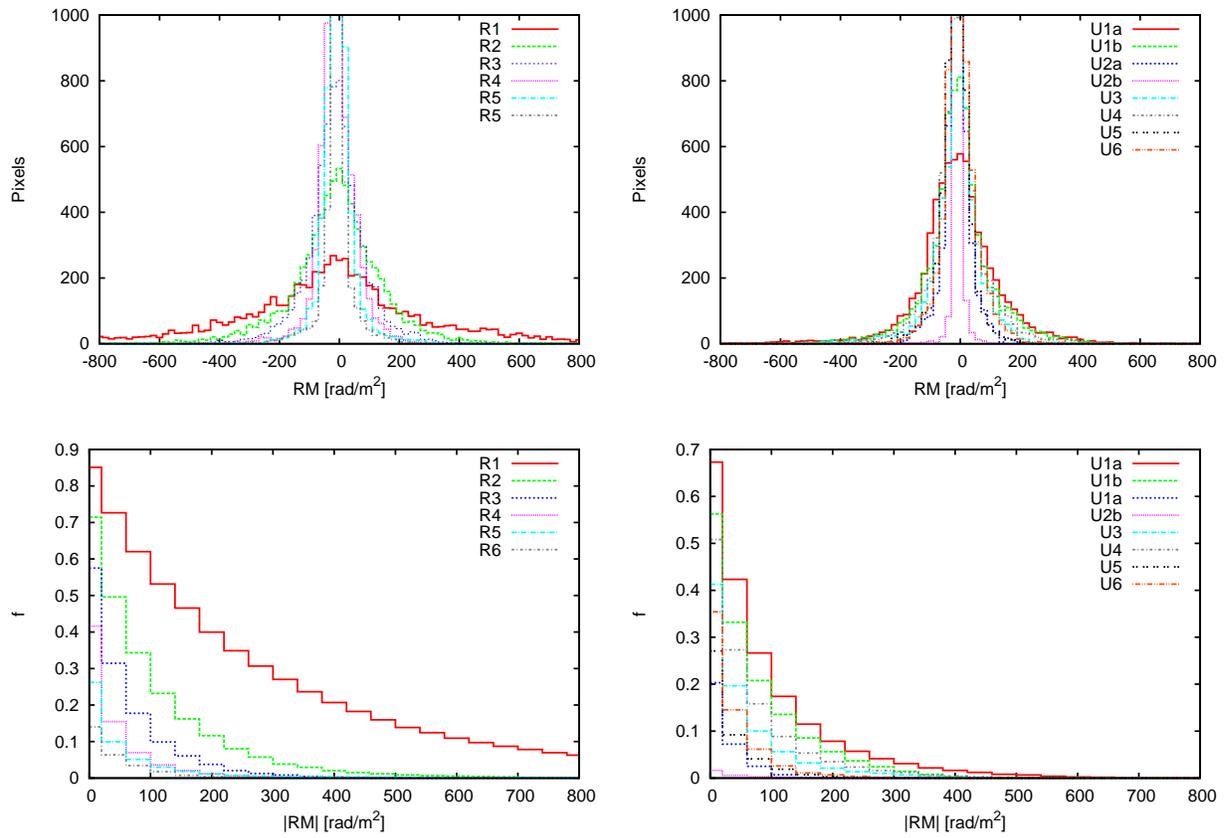,width=1.0\textwidth}
\end{center}
\caption{
Histogram of RM (top) and complementary cumulative histogram of $|RM|$ (bottom) of the central circle of 0.5 Mpc radius of the RM maps shown in Figure \ref{fig:rm}. 
The left panel shows the relaxed clusters, while the right panel shows the unrelaxed clusters.
 \label{fig:rm_histogram}}
\end{figure}

We also plot the area histograms of RM and the complementary cumulative histograms of $|RM|$
inside 500 kpc central circles of the clusters in Figure \ref{fig:rm_histogram}.  Relaxed clusters generally
have more areas covered by significant $|RM|$. For all 
cases, the distribution of positive and negative RM is roughly symmetric.
This is because there is no net magnetic fluxes in the clusters, for the model of field injection 
we have used.  The $RM$  area histograms are similar for clusters of similar sizes and dynamical states,  
though their RM maps may look quite different from their different formation or magnetism history. 
This suggests that the histograms of RM distribution are not sensitive to 
the cluster merger and magnetism history, which can be reflected by their RM filament and patchy 
structures.

\section{Conclusions}

In this paper, we report an ensemble of simulations of magnetic field evolution in galaxy clusters
with a wide range of masses between 1 $\times$ 10$^{14}$ and 2 $\times$ 10$^{15}$ M$_\odot$ in various dynamical
states.  With similar amounts of initial magnetic fields injected from a single AGN at a high redshift, all clusters are eventually filled with 
micro Gauss magnetic fields, except for dynamically very young clusters. The power spectra of 
kinetic energy density show that the ICM is in a turbulent state, while the spectra of magnetic 
energy density show that the small-scale dynamo process is being driven by 
by the ICM turbulence. This result, along with the previous 
study of a single cluster with the magnetic field injections of different amounts of magnetic energy 
and at different redshifts \citep{Xu10}, suggests that magnetization of galaxy clusters
by the operation of small-scale dynamo with the seed magnetic fields from AGN is very robust,
and it produces magnetic fields consistent with observed magnetic fields.  

The magnetic field evolution and distribution are decided both by the masses of the clusters and their dynamical formation
histories. The total magnetic energy is approximately scaled as the square of the virial mass of the cluster, while the dynamically older (relaxed) clusters 
usually have more magnetic fields. This implies that the cluster magnetic fields are mostly determined by the dynamo process of the ICM turbulence
generated by the hierarchical mergers.  Additional 
magnetic fields from more AGNs or smaller cluster forming halos will not have a major impact on the final 
magnetic fields in a massive cluster. 
The $\gamma$ in the scaling relation between magnetic field and gas density radial profiles ($|B|$ $\propto$ n$^{\gamma}$)
range between 0.3 and 0.81 for our simulated clusters, while relaxed clusters have flatter magnetic fields profiles.  
In addition, the relaxed clusters usually have self-similar magnetic field radial distribution, while the field distribution in younger 
unrelaxed clusters is disturbed by their recently mergers.  Though our simulated 
clusters only have initial magnetic fields from a single local source, most volumes 
in the simulated clusters are well magnetized.

We also studied Faraday rotation measurements obtained from the magnetic field and gas density distribution in our simulated clusters. 
They are also determined by both the cluster sizes and their dynamical states. The radial profiles of $|RM|$ resemble
the profiles of the magnetic field strength, and are consistent with the pattern from the observational data. 
The RM maps reflect the recent cluster mergers, as well as the cluster magnetism history.  There are very long 
filaments in the RM maps in the recently merged clusters, while their small scale patchy bands reflect the ICM turbulence.  
Very different distributions of RM of the same cluster but with magnetic fields injected in different locations 
are observed.  This suggests that RM distribution may be a good probe not only for the cluster formation but also its
magnetism history. A detailed study on the relation of RM from simulated clusters
with their ICM turbulence will be presented in a forthcoming paper.

\acknowledgments
  This work was supported by the LDRD and IGPP programs at LANL and by DOE office of science via CMSO. Computations
  were performed using the institutional computing resources at LANL.
  ENZO$\_$MHD is developed at the Laboratory for Computational Astrophysics,
  UCSD with partial support from NSF AST-0808184 to M.L.N.

\bibliographystyle{apj}

\begin{thebibliography}{53}
\expandafter\ifx\csname natexlab\endcsname\relax\def\natexlab#1{#1}\fi

\bibitem[{{Bonafede} {et~al.}(2010){Bonafede}, {Feretti}, {Murgia}, {Govoni},
  {Giovannini}, {Dallacasa}, {Dolag}, \& {Taylor}}]{Bonafede10}
{Bonafede}, A., {Feretti}, L., {Murgia}, M., {Govoni}, F., {Giovannini}, G.,
  {Dallacasa}, D., {Dolag}, K., \& {Taylor}, G.~B. 2010, \aap, 513, A30+

\bibitem[{{Brandenburg} \& {Subramanian}(2005)}]{Brandenburg05}
{Brandenburg}, A., \& {Subramanian}, K. 2005, \physrep, 417, 1

\bibitem[{{Bryan} \& {Norman}(1998)}]{Bryan98}
{Bryan}, G.~L., \& {Norman}, M.~L. 1998, \apj, 495, 80

\bibitem[{{Burbidge}(1959)}]{Burbidge59}
{Burbidge}, G.~R. 1959, \apj, 129, 849

\bibitem[{{Carilli} \& {Taylor}(2002)}]{Carilli02}
{Carilli}, C.~L., \& {Taylor}, G.~B. 2002, \araa, 40, 319

\bibitem[{{Clarke} {et~al.}(2001){Clarke}, {Kronberg}, \&
  {B{\"o}hringer}}]{Clarke01}
{Clarke}, T.~E., {Kronberg}, P.~P., \& {B{\"o}hringer}, H. 2001, \apjl, 547,
  L111

\bibitem[{{Colgate} \& {Li}(2000)}]{Colgate00}
{Colgate}, S.~A., \& {Li}, H. 2000, in IAU Symposium, Vol. 195, Highly
  Energetic Physical Processes and Mechanisms for Emission from Astrophysical
  Plasmas, ed. P.~C.~H. {Martens}, S.~{Tsuruta}, \& M.~A. {Weber}, 255--264

\bibitem[{{Collins} {et~al.}(2010){Collins}, {Xu}, {Norman}, {Li}, \&
  {Li}}]{Collins09}
{Collins}, D.~C., {Xu}, H., {Norman}, M.~L., {Li}, H., \& {Li}, S. 2010, \apjs,
  186, 308

\bibitem[{{Croston} {et~al.}(2005){Croston}, {Hardcastle}, {Harris}, {Belsole},
  {Birkinshaw}, \& {Worrall}}]{Croston05}
{Croston}, J.~H., {Hardcastle}, M.~J., {Harris}, D.~E., {Belsole}, E.,
  {Birkinshaw}, M., \& {Worrall}, D.~M. 2005, \apj, 626, 733

\bibitem[{{Dolag} {et~al.}(2002){Dolag}, {Bartelmann}, \& {Lesch}}]{Dolag02}
{Dolag}, K., {Bartelmann}, M., \& {Lesch}, H. 2002, \aap, 387, 383

\bibitem[{{Dolag} {et~al.}(2001){Dolag}, {Schindler}, {Govoni}, \&
  {Feretti}}]{Dolag01}
{Dolag}, K., {Schindler}, S., {Govoni}, F., \& {Feretti}, L. 2001, \aap, 378,
  777

\bibitem[{{Donnert} {et~al.}(2009){Donnert}, {Dolag}, {Lesch}, \&
  {M{\"u}ller}}]{Donnert08}
{Donnert}, J., {Dolag}, K., {Lesch}, H., \& {M{\"u}ller}, E. 2009, \mnras, 392,
  1008

\bibitem[{{Dubois} {et~al.}(2009){Dubois}, {Devriendt}, {Slyz}, \&
  {Silk}}]{Dubois09}
{Dubois}, Y., {Devriendt}, J., {Slyz}, A., \& {Silk}, J. 2009, \mnras, 399, L49

\bibitem[{{Dubois} \& {Teyssier}(2008)}]{Dubois08}
{Dubois}, Y., \& {Teyssier}, R. 2008, \aap, 482, L13

\bibitem[{{Eilek} \& {Owen}(2002)}]{Eilek02}
{Eilek}, J.~A., \& {Owen}, F.~N. 2002, \apj, 567, 202

\bibitem[{{Eisenstein} \& {Hu}(1999)}]{Eisenstein99}
{Eisenstein}, D.~J., \& {Hu}, W. 1999, \apj, 511, 5

\bibitem[{{En{\ss}lin} \& {Vogt}(2006)}]{Ensslin06}
{En{\ss}lin}, T.~A., \& {Vogt}, C. 2006, \aap, 453, 447

\bibitem[{{Federrath} {et~al.}(2011){Federrath}, {Sur}, {Schleicher},
  {Banerjee}, \& {Klessen}}]{Federrath11}
{Federrath}, C., {Sur}, S., {Schleicher}, D.~R.~G., {Banerjee}, R., \&
  {Klessen}, R.~S. 2011, \apj, 731, 62

\bibitem[{{Feretti}(1999)}]{Feretti99}
{Feretti}, L. 1999, in Diffuse Thermal and Relativistic Plasma in Galaxy
  Clusters, ed. H.~{Boehringer}, L.~{Feretti}, \& P.~{Schuecker}, 3--8

\bibitem[{{Ferrari} {et~al.}(2008){Ferrari}, {Govoni}, {Schindler}, {Bykov}, \&
  {Rephaeli}}]{Ferrari08}
{Ferrari}, C., {Govoni}, F., {Schindler}, S., {Bykov}, A.~M., \& {Rephaeli}, Y.
  2008, Space Sci. Rev., 134, 93

\bibitem[{{Furlanetto} \& {Loeb}(2001)}]{Furlanetto01}
{Furlanetto}, S.~R., \& {Loeb}, A. 2001, \apj, 556, 619

\bibitem[{{Giovannini} {et~al.}(2009){Giovannini}, {Bonafede}, {Feretti},
  {Govoni}, {Murgia}, {Ferrari}, \& {Monti}}]{Giovanini09}
{Giovannini}, G., {Bonafede}, A., {Feretti}, L., {Govoni}, F., {Murgia}, M.,
  {Ferrari}, F., \& {Monti}, G. 2009, \aap, 507, 1257

\bibitem[{{Govoni} {et~al.}(2010){Govoni}, {Dolag}, {Murgia}, {Feretti},
  {Schindler}, {Giovannini}, {Boschin}, {Vacca}, \& {Bonafede}}]{Govoni10}
{Govoni}, F., {Dolag}, K., {Murgia}, M., {Feretti}, L., {Schindler}, S.,
  {Giovannini}, G., {Boschin}, W., {Vacca}, V., \& {Bonafede}, A. 2010, \aap,
  522, A105+

\bibitem[{{Govoni} {et~al.}(2006){Govoni}, {Murgia}, {Feretti}, {Giovannini},
  {Dolag}, \& {Taylor}}]{Govoni06}
{Govoni}, F., {Murgia}, M., {Feretti}, L., {Giovannini}, G., {Dolag}, K., \&
  {Taylor}, G.~B. 2006, \aap, 460, 425

\bibitem[{{Guidetti} {et~al.}(2008){Guidetti}, {Murgia}, {Govoni}, {Parma},
  {Gregorini}, {de Ruiter}, {Cameron}, \& {Fanti}}]{Guidetti08}
{Guidetti}, D., {Murgia}, M., {Govoni}, F., {Parma}, P., {Gregorini}, L., {de
  Ruiter}, H.~R., {Cameron}, R.~A., \& {Fanti}, R. 2008, \aap, 483, 699

\bibitem[{{Kritsuk} {et~al.}(2009){Kritsuk}, {Ustyugov}, {Norman}, \&
  {Padoan}}]{Kritsuk09}
{Kritsuk}, A.~G., {Ustyugov}, S.~D., {Norman}, M.~L., \& {Padoan}, P. 2009, in
  Astronomical Society of the Pacific Conference Series, Vol. 406, Numerical
  Modeling of Space Plasma Flows: ASTRONUM-2008, ed. {N.~V.~Pogorelov,
  E.~Audit, P.~Colella, \& G.~P.~Zank}, 15--+

\bibitem[{{Kronberg} {et~al.}(2001){Kronberg}, {Dufton}, {Li}, \&
  {Colgate}}]{Kronberg01}
{Kronberg}, P.~P., {Dufton}, Q.~W., {Li}, H., \& {Colgate}, S.~A. 2001, \apj,
  560, 178

\bibitem[{{Kulsrud} {et~al.}(1997){Kulsrud}, {Cen}, {Ostriker}, \&
  {Ryu}}]{Kulsrud97}
{Kulsrud}, R.~M., {Cen}, R., {Ostriker}, J.~P., \& {Ryu}, D. 1997, \apj, 480,
  481

\bibitem[{{Li} {et~al.}(2006){Li}, {Lapenta}, {Finn}, {Li}, \&
  {Colgate}}]{Li06}
{Li}, H., {Lapenta}, G., {Finn}, J.~M., {Li}, S., \& {Colgate}, S.~A. 2006,
  \apj, 643, 92

\bibitem[{{McNamara} \& {Nulsen}(2007)}]{McNamara07}
{McNamara}, B.~R., \& {Nulsen}, P.~E.~J. 2007, \araa, 45, 117

\bibitem[{{Navarro} {et~al.}(1995){Navarro}, {Frenk}, \& {White}}]{Navarro95}
{Navarro}, J.~F., {Frenk}, C.~S., \& {White}, S.~D.~M. 1995, \mnras, 275, 720

\bibitem[{{Roettiger} {et~al.}(1999){Roettiger}, {Stone}, \&
  {Burns}}]{Roettiger99}
{Roettiger}, K., {Stone}, J.~M., \& {Burns}, J.~O. 1999, \apj, 518, 594

\bibitem[{{Ryu} {et~al.}(2008){Ryu}, {Kang}, {Cho}, \& {Das}}]{Ryu08}
{Ryu}, D., {Kang}, H., {Cho}, J., \& {Das}, S. 2008, Science, 320, 909

\bibitem[{{Schekochihin} \& {Cowley}(2006)}]{Schekochihin06}
{Schekochihin}, A.~A., \& {Cowley}, S.~C. 2006, Physics of Plasmas, 13, 056501

\bibitem[{{Schekochihin} {et~al.}(2008){Schekochihin}, {Cowley}, {Kulsrud},
  {Rosin}, \& {Heinemann}}]{Schekochihin08}
{Schekochihin}, A.~A., {Cowley}, S.~C., {Kulsrud}, R.~M., {Rosin}, M.~S., \&
  {Heinemann}, T. 2008, Physical Review Letters, 100, 081301

\bibitem[{{Schleicher} {et~al.}(2010){Schleicher}, {Banerjee}, {Sur},
  {Arshakian}, {Klessen}, {Beck}, \& {Spaans}}]{Schleicher10}
{Schleicher}, D.~R.~G., {Banerjee}, R., {Sur}, S., {Arshakian}, T.~G.,
  {Klessen}, R.~S., {Beck}, R., \& {Spaans}, M. 2010, \aap, 522, A115+

\bibitem[{{Spergel} {et~al.}(2007){Spergel}, {Bean}, {Dor{\'e}}, {Nolta},
  {Bennett}, {Dunkley}, {Hinshaw}, {Jarosik}, {Komatsu}, {Page}, {Peiris},
  {Verde}, {Halpern}, {Hill}, {Kogut}, {Limon}, {Meyer}, {Odegard}, {Tucker},
  {Weiland}, {Wollack}, \& {Wright}}]{Spergel07}
{Spergel}, D.~N., {Bean}, R., {Dor{\'e}}, O., {Nolta}, M.~R., {Bennett}, C.~L.,
  {Dunkley}, J., {Hinshaw}, G., {Jarosik}, N., {Komatsu}, E., {Page}, L.,
  {Peiris}, H.~V., {Verde}, L., {Halpern}, M., {Hill}, R.~S., {Kogut}, A.,
  {Limon}, M., {Meyer}, S.~S., {Odegard}, N., {Tucker}, G.~S., {Weiland},
  J.~L., {Wollack}, E., \& {Wright}, E.~L. 2007, \apjs, 170, 377

\bibitem[{{Subramanian} {et~al.}(2006){Subramanian}, {Shukurov}, \&
  {Haugen}}]{Subramanian06}
{Subramanian}, K., {Shukurov}, A., \& {Haugen}, N.~E.~L. 2006, \mnras, 366,
  1437

\bibitem[{{Sunyaev} {et~al.}(2003){Sunyaev}, {Norman}, \& {Bryan}}]{Sunyaev03}
{Sunyaev}, R.~A., {Norman}, M.~L., \& {Bryan}, G.~L. 2003, Astronomy Letters,
  29, 783

\bibitem[{{Sur} {et~al.}(2010){Sur}, {Schleicher}, {Banerjee}, {Federrath}, \&
  {Klessen}}]{Sur10}
{Sur}, S., {Schleicher}, D.~R.~G., {Banerjee}, R., {Federrath}, C., \&
  {Klessen}, R.~S. 2010, \apjl, 721, L134

\bibitem[{{Taylor} \& {Perley}(1993)}]{Taylor93}
{Taylor}, G.~B., \& {Perley}, R.~A. 1993, \apj, 416, 554

\bibitem[{{Turk} {et~al.}(2011){Turk}, {Smith}, {Oishi}, {Skory}, {Skillman},
  {Abel}, \& {Norman}}]{Turk11}
{Turk}, M.~J., {Smith}, B.~D., {Oishi}, J.~S., {Skory}, S., {Skillman}, S.~W.,
  {Abel}, T., \& {Norman}, M.~L. 2011, \apjs, 192, 9

\bibitem[{{van Weeren} {et~al.}(2010){van Weeren}, {R{\"o}ttgering},
  {Br{\"u}ggen}, \& {Hoeft}}]{Weeren10}
{van Weeren}, R.~J., {R{\"o}ttgering}, H.~J.~A., {Br{\"u}ggen}, M., \& {Hoeft},
  M. 2010, Science, 330, 347

\bibitem[{{Vazza} {et~al.}(2011){Vazza}, {Brunetti}, {Gheller}, {Brunino}, \&
  {Br{\"u}ggen}}]{Vazza11}
{Vazza}, F., {Brunetti}, G., {Gheller}, C., {Brunino}, R., \& {Br{\"u}ggen}, M.
  2011, \aap, 529, A17+

\bibitem[{{Vazza} {et~al.}(2009){Vazza}, {Brunetti}, {Kritsuk}, {Wagner},
  {Gheller}, \& {Norman}}]{Vazza09}
{Vazza}, F., {Brunetti}, G., {Kritsuk}, A., {Wagner}, R., {Gheller}, C., \&
  {Norman}, M. 2009, \aap, 504, 33

\bibitem[{{Vazza} {et~al.}(2006){Vazza}, {Tormen}, {Cassano}, {Brunetti}, \&
  {Dolag}}]{Vazza06}
{Vazza}, F., {Tormen}, G., {Cassano}, R., {Brunetti}, G., \& {Dolag}, K. 2006,
  \mnras, 369, L14

\bibitem[{{Vogt} \& {En{\ss}lin}(2005)}]{Vogt05}
{Vogt}, C., \& {En{\ss}lin}, T.~A. 2005, \aap, 434, 67

\bibitem[{{Voit}(2005)}]{Voit05}
{Voit}, G.~M. 2005, Reviews of Modern Physics, 77, 207

\bibitem[{{Widrow}(2002)}]{Widrow02}
{Widrow}, L.~M. 2002, Reviews of Modern Physics, 74, 775

\bibitem[{{Xu}(2009)}]{Xu09b}
{Xu}, H. 2009, PhD thesis, University of California, San Diego

\bibitem[{{Xu} {et~al.}(2008){Xu}, {Li}, {Collins}, {Li}, \& {Norman}}]{Xu08a}
{Xu}, H., {Li}, H., {Collins}, D., {Li}, S., \& {Norman}, M.~L. 2008, \apjl,
  681, L61

\bibitem[{{Xu} {et~al.}(2009){Xu}, {Li}, {Collins}, {Li}, \& {Norman}}]{Xu09}
{Xu}, H., {Li}, H., {Collins}, D.~C., {Li}, S., \& {Norman}, M.~L. 2009, \apjl,
  698, L14

\bibitem[{{Xu} {et~al.}(2010){Xu}, {Li}, {Collins}, {Li}, \& {Norman}}]{Xu10}
---. 2010, \apj, 725, 2152

\end{thebibliography}

\clearpage

\clearpage

\end{document}